\documentclass{aastex}
\usepackage{spr-astr-addons}
\usepackage{url}

\RequirePackage{color}

\begin{document}

\title{The study of s-process nucleosynthesis based on barium stars, CEMP-s and CEMP-r/s stars}
\slugcomment{Not to appear in Nonlearned J., 45.}
\shorttitle{The study of s-process nucleosynthesis based on peculiar stars}
\shortauthors{Cui et al.}

\author{Cui Wenyuan\altaffilmark{1,2,3}}
\affil{e-mail: cuiwenyuan@hebtu.edu.cn}
\and \author{Shi Jianrong\altaffilmark{2}}
\and \author{Geng Yuanyuan\altaffilmark{1}}
\and \author{Zhang Caixia\altaffilmark{1}}
\and \author{Meng Xiaoying\altaffilmark{1}}
\and \author{Shao Lang\altaffilmark{1}}
\author{Zhang Bo\altaffilmark{1}*}
\affil{zhangbo@hebtu.edu.cn}

\altaffiltext{1}{Department of Physics, Hebei Normal University, 20 Nanerhuan
Dong Road, Yuhua District, Shijiazhuang 050024, P.R.China.}
\altaffiltext{2}{National Astronomical observatories, Chinese Academy
of Sciences, 20A Datun Road, Chaoyang District, Beijing 100012,
P.R.China.}
\altaffiltext{3}{Department of Physics, Shandong University, Jinan, Shandong 250100,
P.R.China.}

\begin{abstract}
In order to get a broader view of the s-process nucleosynthesis we
study the abundance distribution of heavy elements of 35 barium
stars and 24 CEMP-stars, including nine CEMP-s stars and 15
CEMP-r/s stars. The similar distribution of [Pb/hs] between CEMP-s
and CEMP-r/s stars indicate that the s-process material of both
CEMP-s and CEMP-r/s stars should have a uniform origin, i.e. mass
transfer from their predominant AGB companions. For the CEMP-r/s
stars, we found that the r-process should provide similar proportional
contributes to the second s-peak and the third s-peak elements,
and also be responsible for the higher overabundance of
heavy elements than those in CEMP-s stars. Which hints that the
r-process origin of CEMP-r/s stars should be closely linked to the
main r-process. The fact that some small $r$ values exist for both
barium and CEMP-s stars, implies that the single exposure event of
the s-process nucleosynthesis should be general in a wide
metallicity range of our Galaxy. Based on the relation between
$C_{r}$ and $C_{s}$, we suggest that the origin of r-elements for
CEMP-r/s stars have more sources. A common scenario is that the
formation of the binary system was triggered by only one or a few
supernova. In addition, accretion-induced collapse(AIC) or SN 1.5
should be the supplementary scenario, especially for these whose pre-AGB
companion with higher mass and smaller orbit radius, which support
the higher values of both $C_{r}$ and $C_{s}$.
\end{abstract}



\section{Introduction}

Based on whether the timescale for neutron capture is slower or faster
than the $\beta$-decay timescale for unstable nuclei, the nuclei beyond
the iron group are created in neutron-capture processes, i.e. s- (slow)
or r- (rapid). The two neutron-capture processes are thought to  occur
under different physical conditions and therefore likely to arise in
different astrophysical sites. The s-process, which requires a lower
neutron flux (with a typical neutron-capture taking many years), is
generally thought to occur during the double-shell burning phase of
asymptotic giant branch (AGB) stars with low- and intermediate-mass
\citep{bus99}. The r-process requires a high neutron flux level
(with many neutron-captures over a timescale of a fraction of a
second), which is expected to take place in the exploding astrophysical
site or sites such as the $\nu$-driven wind of Type II (i.e. core-collapse)
supernovae \citep{woo94}, the mergers of neutron stars \citep{ros00},
accretion-induced collapse \citep[AIC;][]{qia03}, and Type 1.5 supernovae \citep{zij04} etc.

Low-mass AGB stars are as well known as the main site for the s-process
elements  producing \citep{gal98,lug03,her04,kap11,bis11}. During their
thermally pulsing (TP) phase of AGB stars, some protons are assumed to
penetrate into the top layers of the He-intershell from the bottom of
the convective envelope, and then captured by the freshly synthesised
$^{12}$C then directly producing $^{13}$C via the $^{12}$C(p, $\gamma$)
$^{13}$N($\beta^{+}$, $\nu$)$^{13}$C reactions \citep{ibe83}.
The primary-like reaction, $^{13}$C($\alpha$, n)$^{16}$O which works
at a temperature of about $0.1\times10^{9}$ K, is generally regarded
as the major neutron source of the s-process nucleosynthesis. Based
on a specific hydrodynamical treatment, \citet{sne08} pointed out
that at the border between the upper H-rich convective envelope and
the lower C-rich and He-rich radiative He-intershell, some physical
mechanisms such as rotational mixing, shear turbulence, gravitational
waves, thermodynamical instability and so on could affect the downflow
of protons from the envelope. In fact, the precise mechanism for how
protons penetrate into the He-intershell is still unknown. Finally,
C and s-elements were brought to the surface of AGB stars from He-intershell
by the recurrent third dredge-up (TDU) events. Then, different degrees of
abundance enhancement of these elements should be observed, which is
related to different initial masses and metallicities.

So far, many metal-poor stars have been found based on some special
surveys for metal-poor stars which finished or undergoing now, such
as the HK survey \citep{bee92,bee07}, the Hamburg/ESO Survey
\citep{chr03}, the Sloan Digital Sky Survey (SDSS) \citep{yor00}, as
well as the subprogram of SDSS, i.e. the Sloan Extension for
Galactic Understanding and Exploration \citep[SEGUE, ][]{yan09}.
Where, at least $10\%$ and probably as much as $21\%$ of Galactic
stars with [Fe/H]$\leq-2.0$ were identified as carbon-enhanced
metal-poor (CEMP, with [C/Fe]$>1.0$) stars \citep{luc06,fre07}.
\citet{bee05} have classified CEMP-stars as CEMP-s, CEMP-r, CEMP-r/s
and CEMP no in which no enhanced r- or s-elements observed. These
old stars with low initial mass ($M<0.9 M_{\odot}$) are still on
their main-sequence or giant evolution phase. \citet{aok07} reported
that about $80\%$ of CEMP-stars are CEMP-s stars which showing high
enhancement of s-elements. The plausible scenario for the origin of
the C and s-elements of CEMP-stars is usually attributed to mass
transfer from their former AGB companions (white dwarfs now) by the
stellar wind-accretion in binary systems. It is as well known that
Eu is one of the most readily measurable elements in optical spectra
of metal-poor stars, which is mainly synthesised through r-process.
Actually, Eu is always used as a represent element of r-process. At
present, about half of CEMP-s stars are found with higher Eu
enhancement, i.e. CEMP-r/s stars \citep{sne08,kap11,bis11}.

The discovery of CEMP-r/s stars \citep{hil00,coh03} is puzzling,
as their formation requires pollution from both an AGB star and a
supernova. The origin of the abundance peculiarities of CEMP-r/s
stars is sill a debate issue now, and many scenarios have been
presented \citep[see][for detail]{jon06}. \citet{qia03} proposed a
scenario for the formation of CEMP-r/s stars. In this case, the C
and s-process material was firstly produced and then accreted from
an AGB primary star, which has evolved to a white dwarf.
Then, the white dwarf accretes matter from the secondary one
and soon the mass transfer triggered an AIC event.
The neutrino wind created by the collapse produces r-process elements,
which pollutes the secondary one again. However, it is still far from certain
that whether the r-process elements can be produced during AIC.
Another possible scenario is that the AGB star transfers s-rich
materials to the observed star but does not suffer a large mass
loss and at the end of the AGB phase, the degenerate core may
reach the Chandrasekhar mass which leading to a Type 1.5
supernova. The reason is due to the low mass-loss rate of AGB star
at low metallicity \citep{zij04}. Such supernovae could explain
both the enhancement pattern and the metallicity dependence of the
double-enhanced halo stars. The r-process prepollution is also a
possible scenario for CEMP-r/s star formation. In this picture,
the formation of binary systems was triggered by a supernova which
polluted and clumped a nearby molecular cloud. So, the observed
star was firstly enhanced with r-process elements as it once
formed, and then received large amounts of s-process elements from
its massive AGB companion and finally turned into a CEMP-r/s star
\citep{del04,bar05,gal05,iva05}. \citet{aok02} proposed that this
scenario could possibly explain the formation of CEMP-r/s stars.

Excepting CEMP-s stars, some peculiar red giants with high
metallicity similar to the Sun are known as ``barium stars'', or
``BaII stars'' which were firstly defined by \citet{bid51}.
Initially, barium stars only included G and K giants which exhibit
enhanced features of BaII, SrII, CH, CN lines and sometimes
C$_{2}$ molecular bands. Many qualitative studies on barium stars
have been developed \citep [e.g.][]
{bur57,dan65,pil77,smi84,zac94,lia03,all06a,all06b,smi07,hus09}.
Barium stars could not be self-enriched, i.e. synthesizing
s-process elements in interiors and dredging s-rich materials up
to the surface of envelope, because of their lower mass and lower
luminosity than those of AGB stars. Periodic variations of the
radial velocity have been detected for many barium stars,
supporting the scenario that these stars belong to binary systems,
and have accreted s-process material from their AGB companions (a
white dwarf now) \citep{mcc84,mcc90}. \citet{dom83} and
\citet{boh85} have detected some of their white dwarf companions
in the UV with the International Ultraviolet Explorer. Obviously,
both barium stars and CEMP-s stars have the similar origin for
their enhanced s-elements.

With the help of analysing the abundance patterns of heavy
elements enriched stars, the detailed information of s-process
nucleosynthesis which taking place in AGB stars can be obtained.
In this paper, we focus on the s-process nucleosynthesis at
different metallicities, and compare the theoretical yields
calculated by our parametric model \citep{zha06,cui10} with the
sample stars. In order to get a broader view about the s-process
nucleosynthesis, CEMP-s, CEMP-r/s and barium stars were all
included into the sample stars of this work. Moreover, study on
the proper sample stars might enable us to directly investigate
the products of individual processes, s- and r-process here, and
to identify the respective astrophysical sites hoping for
r-process. The chosen sample stars are described in Sect. 2,
the analysing for the abundance ratios distribution of heavy
elements in Sect. 3, and the introduction of the parametric model
in Sect. 4. In Sect. 5 the simulation results and the
nucleosynthetic effects are discussed, while conclusions are drawn
in Sect. 6.

\section{Sample Stars}
It is well known that the abundances of heavy elements, such as
Sr, Ba, Pb, and so on, are enhanced in barium stars
(``metal-rich''), CEMP-s and CEMP-r/s stars, which degree just
differ by metallicity. Recently, more detailed and more precise
abundance data for heavy elements of barium stars have been
presented by some works, e.g. \citet{all06a,pom08}. In addition,
the largest wide-field spectroscopic surveys for metal-poor stars
to date and their following high-resolution observations, e.g. the
HK survey \citep{bee92} and the HES survey \citep{chr03}, have
provided abundant abundance data with high quality for CEMP-s and
CEMP-r/s stars from our early Galaxy. All of these excellent works
make it possible to study the s-process nucleosynthesis at
different metallicities by analysing the abundance pattern of
heavy elements of sample stars with a wide metallicity range.

It is very important to study such kinds of stars for a better
understanding of the efficiency of the s-process at
different metallicity in our Galaxy. We explore the pattern of
s-process nucleosynthesis with metallicity by comparing the
observed abundance of neutron-capture elements with those of the
predicted s- and r-process contributions, as the more complete
abundance pattern for the neutron-capture elements provides an
improved constraint on the nature of the s-process
nucleosynthesis. Among barium stars, CEMP-s and CEMP-r/s stars
from the literature, we selected only those with all of the
following four elements detection, i.e. Sr (or Y for the first
s-peak), Ba (for the second s-peak), Eu (for the r-process) and Pb
(for the third s-peak). In order to include sample stars as many
as possible, some stars with one of the four elements mentioned
above, whose upper limit abundance was detected, were also
selected. According to this standard, we collected 27 barium stars
as displayed in table 1, nine CEMP-s and 15
CEMP-r/s stars as displayed in table 3. In addition, eight barium
stars with theoretical values of Pb abundance predicted by
\citet{hus09} were also included, as there are no observational
Pb data available from the literatures.

Spectroscopic data of barium stars were adopted from
\citet{all06a,pom08}, and \citet{smi07}, while for the metal-poor
stars, the observation data were taken from
\citet{aok01,aok02,joh02,luc03,bar05,iva05,jon06,coh03,coh06,gos06,mas06}.

\section{The Abundance Ratio Distribution of heavy elements}
Elements with $Z>30$ are labeled in neutron-capture. Though the
cumulated content of neutron-capture elements is only about
$10^{-6}$ percent by number of Solar material, but they can produce
significant spectral absorption lines, and thus can be detected in
stars over a wide metallicity range \citep{sne08}. So far, many
heavy elements have been observed for stars with different
metallicity. There are three peaks on the s-process path, i.e. the
first s-peak at Sr, Y, Zr (light-s elements, at the neutron magic
number N =50), the second s-peak at Ba, La, Ce, etc (heavy-s
elements, at N =82) and the third peak at Pb and Bi (N =126), which
locate at the termination point of the s-process path. For
convenience, we adopt Ba, La, Ce to represent the heavy-s elements,
which is labeled as ``hs" hereafter, and Sr, Y, Zr to represent the
light-s elements labeled as ``ls".

\begin{figure*}
 \centering
 \includegraphics[width=.60\textwidth,height=.40\textheight]{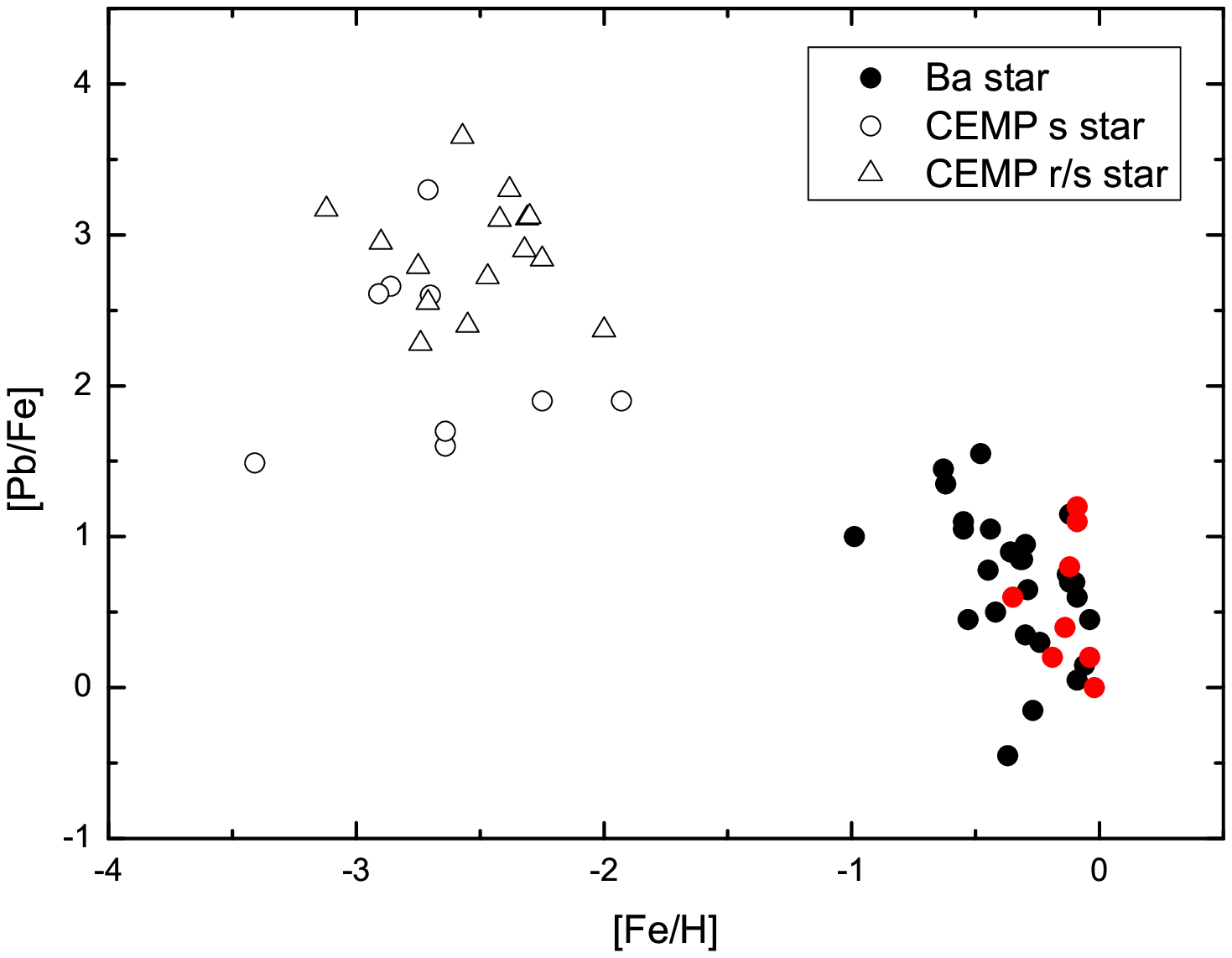}
 \hspace{-4.0cm}
 \includegraphics[width=.60\textwidth,height=.40\textheight]{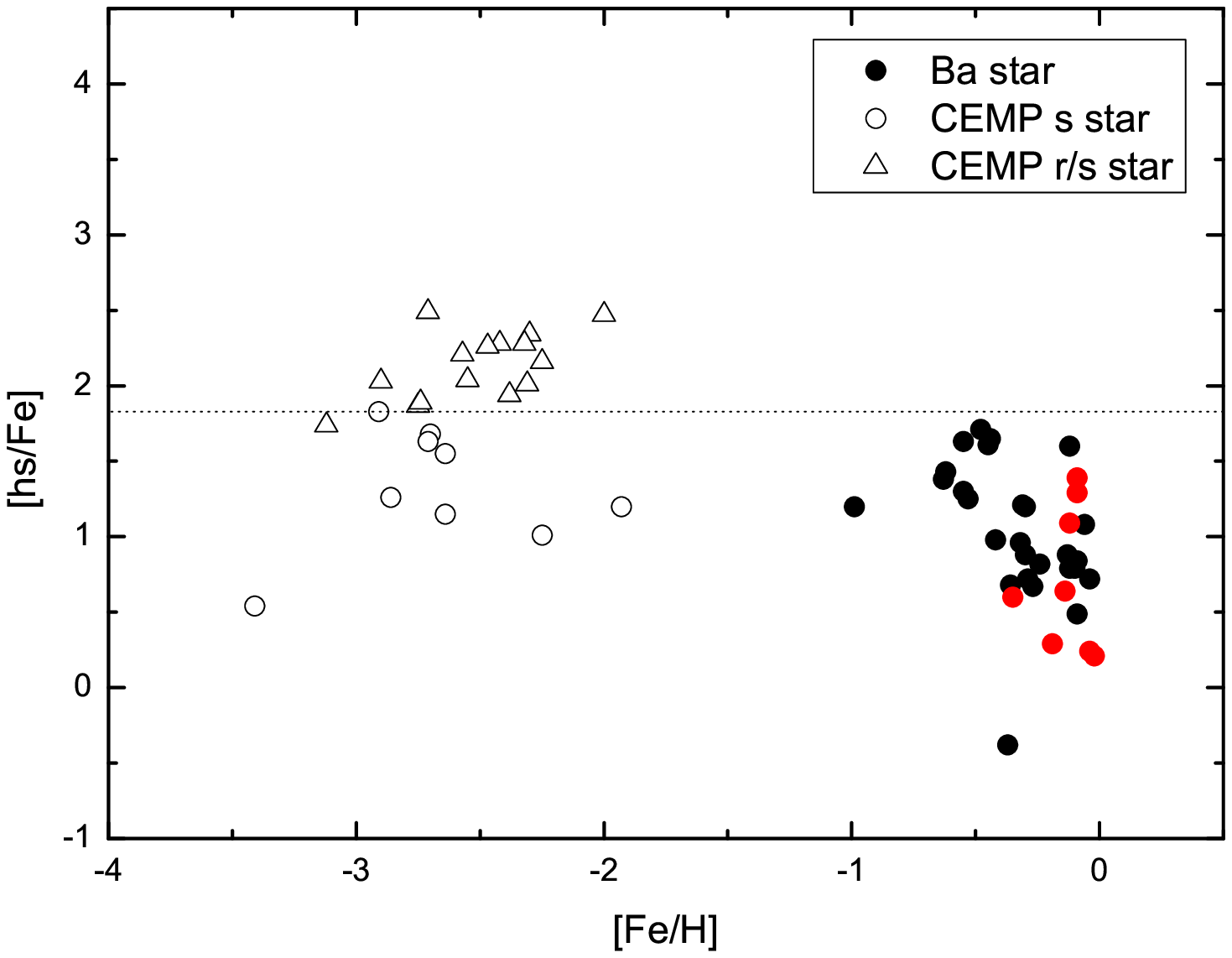}
 \vspace{-3.2cm}
  \caption{(a) [Pb/Fe] vs. [Fe/H]; (b) [hs/Fe] vs. [Fe/H]. The filled
  circles represent barium stars, where the red ones represent the one
whose lead abundance was predicted by \citet{hus09}. The unfilled
circles represent CEMP-s stars, and unfilled triangles represent CEMP-r/s
stars.\label{Fig:XFe}}
 \end{figure*}

In figure~\ref{Fig:XFe}, we show, respectively, [Pb/Fe] and [hs/Fe] as a
function of [Fe/H] with different symbols employed for barium,
CEMP-s, and CEMP-r/s stars. The filled circles represent barium
stars, and the red ones represent the one whose lead abundance was
predicted by \citet{hus09}. The unfilled circles represent CEMP-s
stars, while unfilled triangles represent CEMP-r/s stars. From
figure~\ref{Fig:XFe}, we can see that both [Pb/Fe] and [hs/Fe] have the same
tendency, i.e. increasing with decreasing metallicity, which is
consistent with the s-process calculation results of AGB stars
with low metallicities \citep{gal98,gor00}. For the CEMP samples,
however, [Pb/Fe] locates in the range of $1.5-3.8$ which is almost
higher than that of [hs/Fe] ($\sim$ $0.5-2.5$). The standard AGB
models, which based on the primary like character of the main
neutron source, $^{13}$C($\alpha$, n)$^{16}$O,
predict that low-metallicity AGB stars should exhibit higher
over-abundances of heavier s-elements (Z $\geq$ 56) compared to
the lighter ones (such as Sr, Y, Zr, etc). Especially the hugely
enhanced Pb and Bi should be attributed to the large number ratios
of neutron to iron seeds in low-metallicity environments
\citep{gor00}.

\citet{sne08} pointed out that the s-elements seen in CEMP-s stars
should be the result of mass transfer from the winds of AGB
companions (the undetected white dwarf now), because nearly all of
them are dwarfs or giants with low luminosities, which are much
fainter than AGB stars. Many studies on CEMP-r/s and barium stars
all agree that the observed overabundance of s-process elements
were accreted from the initially more massive star that underwent
the AGB phase \citep{qia03,bar05,gal05,lia03,all06b,hus09},
although the origin of r-enhanced material for CEMP-r/s stars is
still in debate \citep{jon06}. Considering the effect of mass
transfer in a binary system, it is difficult to constrain the
initial abundances of the s-process synthesis regions from
observations of [X/Fe] ratios. Nevertheless, when we study the
s-process nucleosynthesis in AGB stars, the mass transfer
uncertainty could be canceled if using the abundance ratios, e.g.
[Pb/hs] \citep{str06,mas10}. As the intrinsic indices [Pb/hs] are
independent of that whether the s-enhanced star is an intrinsic or
extrinsic AGB, it is very useful to investigate the efficiency of
the s-process.

\begin{figure*}
  \centering
 \includegraphics[width=.60\textwidth,height=.40\textheight]{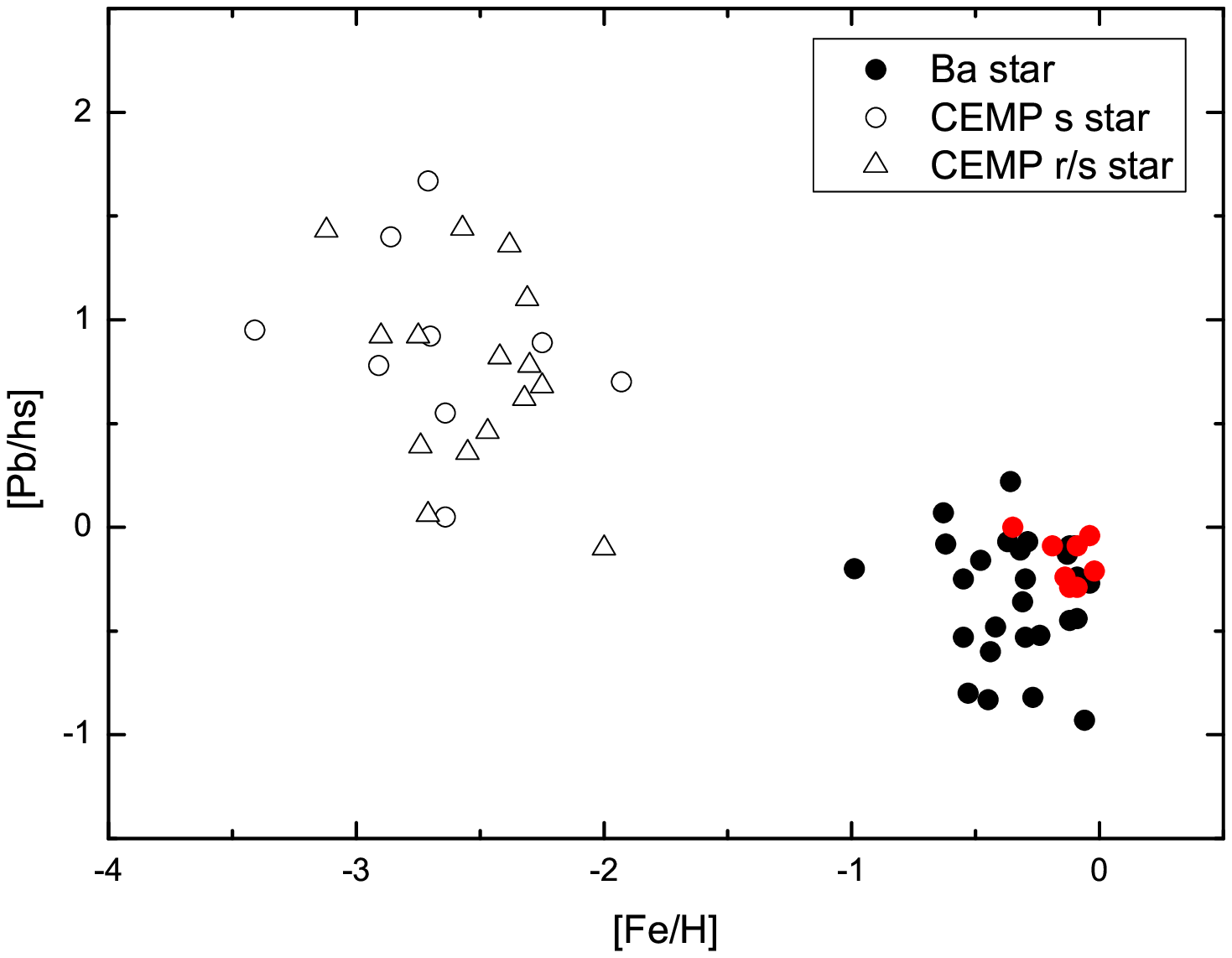}
 \hspace{-4.0cm}
 \includegraphics[width=.60\textwidth,height=.40\textheight]{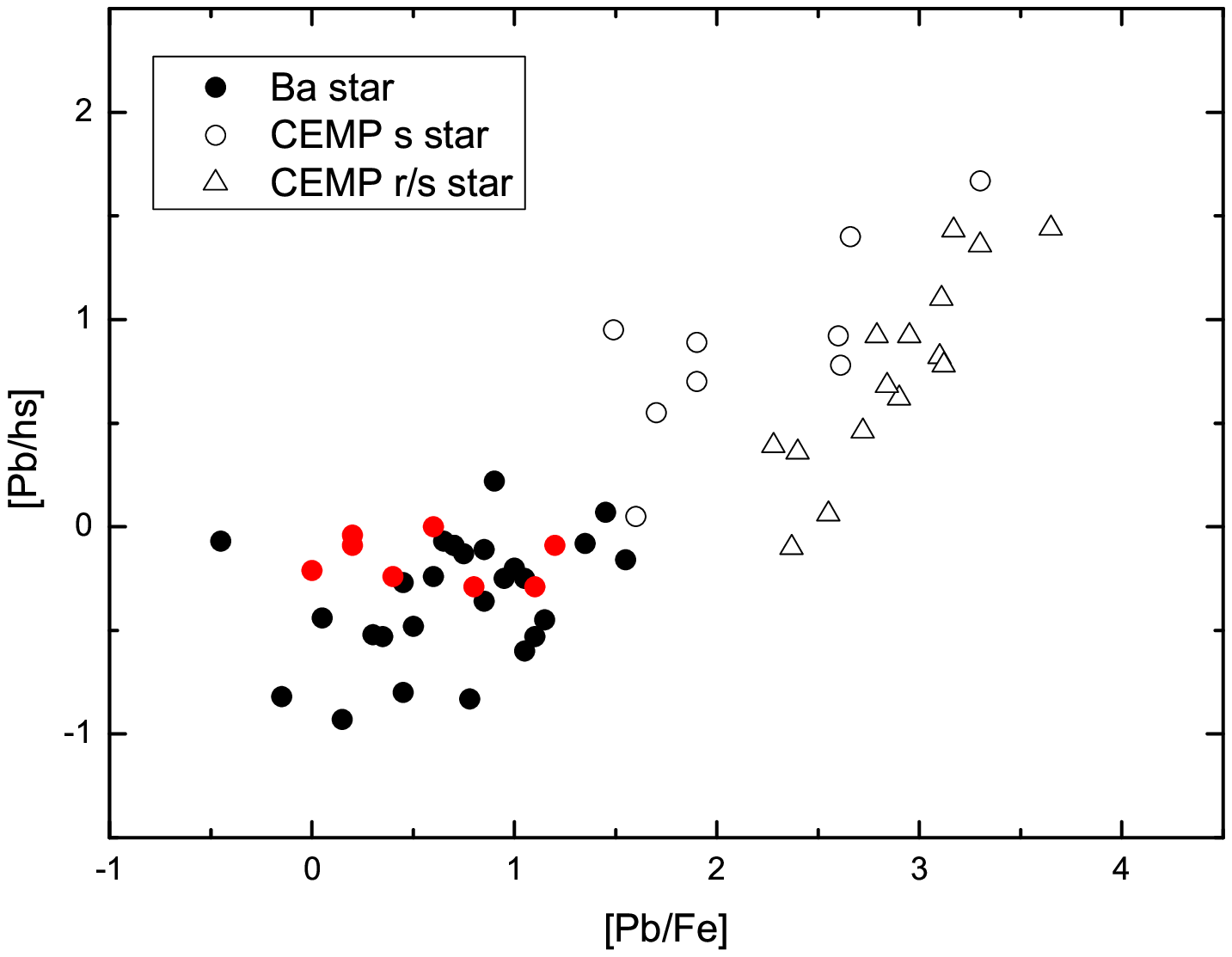}
 \vspace{-3.2cm}
  \caption{(a) [Pb/hs] vs. [Fe/H]; (b) [Pb/hs] vs. [Pb/Fe]. The meaning
  of symbols are same in figure~\ref{Fig:XFe}. \label{Fig:Pbhs}}
\end{figure*}

In figure~\ref{Fig:Pbhs}a,b, we show [Pb/hs] versus [Fe/H] and [Pb/Fe],
respectively, the symbols have the same meanings as in figure~\ref{Fig:XFe}.
We can see that the [Pb/hs] values of both CEMP-s and CEMP-r/s
stars are almost higher than 0, while, on the contrary, they are
almost lower than 0 for barium stars. Moreover, the larger scatter
of [Pb/hs] at low-metallicity exists as discussed by \citet{bus01}
and \citet{kap11}, which are about 2 dex among CEMP-s and CEMP-r/s
stars, while 1 dex in barium stars. In addition, although the
[hs/Fe] of CEMP-r/s stars is almost higher than that of CEMP-s
stars, there are no obviously different in the distribution of
[Pb/hs] between these two type stars, which is consistent with the
result of \citet{bis11}. This result proves that the s-process
material of both CEMP-s and CEMP-r/s stars should have a uniform
origin, i.e. mass transfer from their pre-dominant AGB companions
in the corresponding binary systems.

It is interesting that the [hs/Fe] of CEMP-r/s stars is higher than
that of CEMP-s stars (see figure~\ref{Fig:XFe}b), and the abundance
ratio [Pb/Fe] shows a similar tendency (see figure~\ref{Fig:Pbhs}b),
i.e. at a fixed [Pb/hs] value, [Pb/Fe] of CEMP-r/s stars is always
larger than that of CEMP-s stars. As it is well known that the
neutron-capture elements beyond the iron group are created by some
combination of the s- and r-process nucleosynthesis, with each
responsible for approximately half of the isotopes. This tendency
implies that, excluding the s-process contribution, the r-process of
CEMP-r/s stars should contribute to both the second s-peak and the
third s-peak elements with similar proportion. Which is also
consistent with the proportions of r-process to these elements for
their solar abundances, i.e. Ba $15\%$, La $25\%$, Ce $19\%$ (second
s-peak elements) and Pb $21\%$ (third s-peak elements)
\citep[adopted from][]{bur00}. In other words, the enhanced
over-abundances of heavier s-elements in CEMP-r/s stars than those
in CEMP-s stars, such as Pb, Ba, La, etc, should be attributed to
r-process nucleosynthesis. Recent studies indicate that the main-r
process is responsible for a pure r-process abundance pattern for
the heavy r-process elements $56\leq Z<83$ \citep{sne03,qia07}.
\citet{mon07} suggested that this pattern is remarkably stable from
star to star, and in excellent agreement with the contribution of
the r-process to the solar abundances. Moreover, the stability of
the observed abundance pattern of r-process, and the good agreement
with the solar system r-process contribution imply that the
r-process events generate a universal abundance distribution, which
hints that the r-process origin of CEMP-r/s stars should be closely
linked to the main r-process, especially for the elements with
$Z\geq56$.

\begin{figure*}
 \centering
 \includegraphics[width=.60\textwidth,height=.40\textheight]{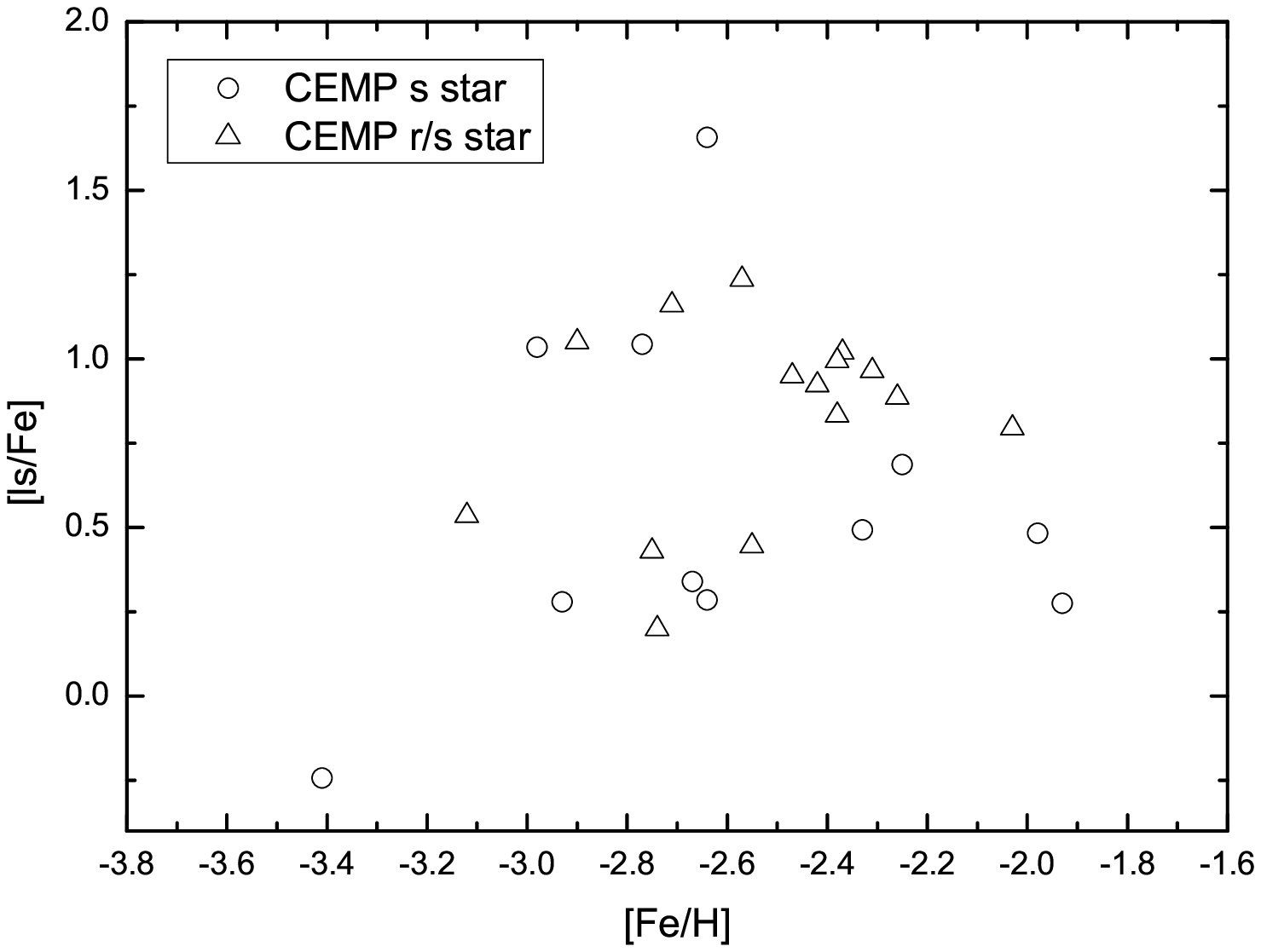}%
 \hspace{-4.0cm}
 \includegraphics[width=.60\textwidth,height=.40\textheight]{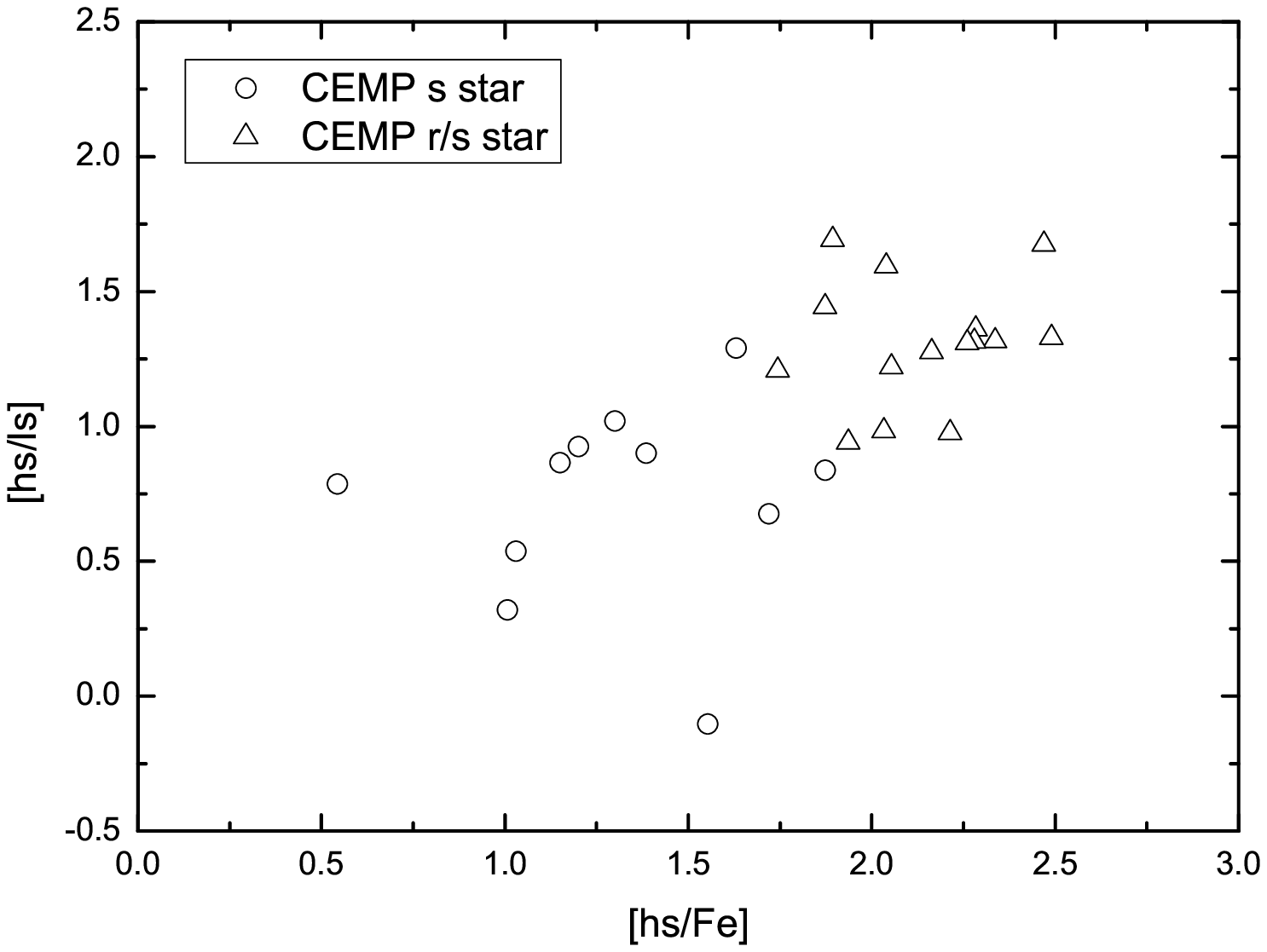}
 \vspace{-3.2cm}
  \caption{(a) [ls/Fe] vs. [Fe/H]; (b) [hs/ls] vs. [hs/Fe]. The meaning
  of symbols are same in figure~\ref{Fig:XFe}. \label{Fig:hsls}}
\end{figure*}

\begin{figure*}
  \vspace{2.2cm}
 \begin{center}
 \includegraphics[width=.60\textwidth,height=.40\textheight]{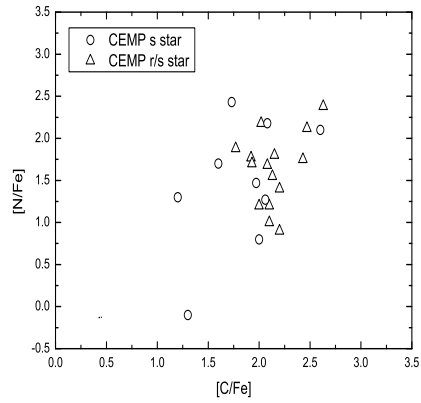}
 \vspace{-3.2cm}
  \caption{[N/Fe] vs. [C/Fe]. The meaning
  of symbols are same in figure~\ref{Fig:XFe}. \label{Fig:cfenfe}}
  \end{center}
\end{figure*}

Figure~\ref{Fig:hsls}a,b show [ls/Fe] versus [Fe/H] and [hs/ls]
versus [hs/Fe], respectively. There is no obvious distribution
difference of [ls/Fe] between CEMP-s and CEMP-r/s stars (see
figure~\ref{Fig:hsls}a). This means that the light-s elements (such
as Sr, Y, Zr, etc.) of such two kinds of stars should have a similar
origin, i.e. mass transfer from their pre-AGB companions. However,
the CEMP-r/s stars almost show higher values of both [hs/ls] and
[hs/Fe] than CEMP-s stars. This could be explained by the fact that
for the CEMP-r/s stars the r-process have done a large contribution
to their heavy-s elements, but a marginal contribution to their
light-s elements. In figure~\ref{Fig:cfenfe}, we present [N/Fe]
versus [C/Fe] for CEMP-s and -r/s stars. The similar distribution of
[N/Fe] and [C/Fe] between CEMP-s and -r/s stars also supports that
they have suffered an AGB pollution which brought much N, C and
s-elements.

\begin{figure*}
 \vspace{2.2cm}
 \centering
 \includegraphics[width=.60\textwidth,height=.40\textheight]{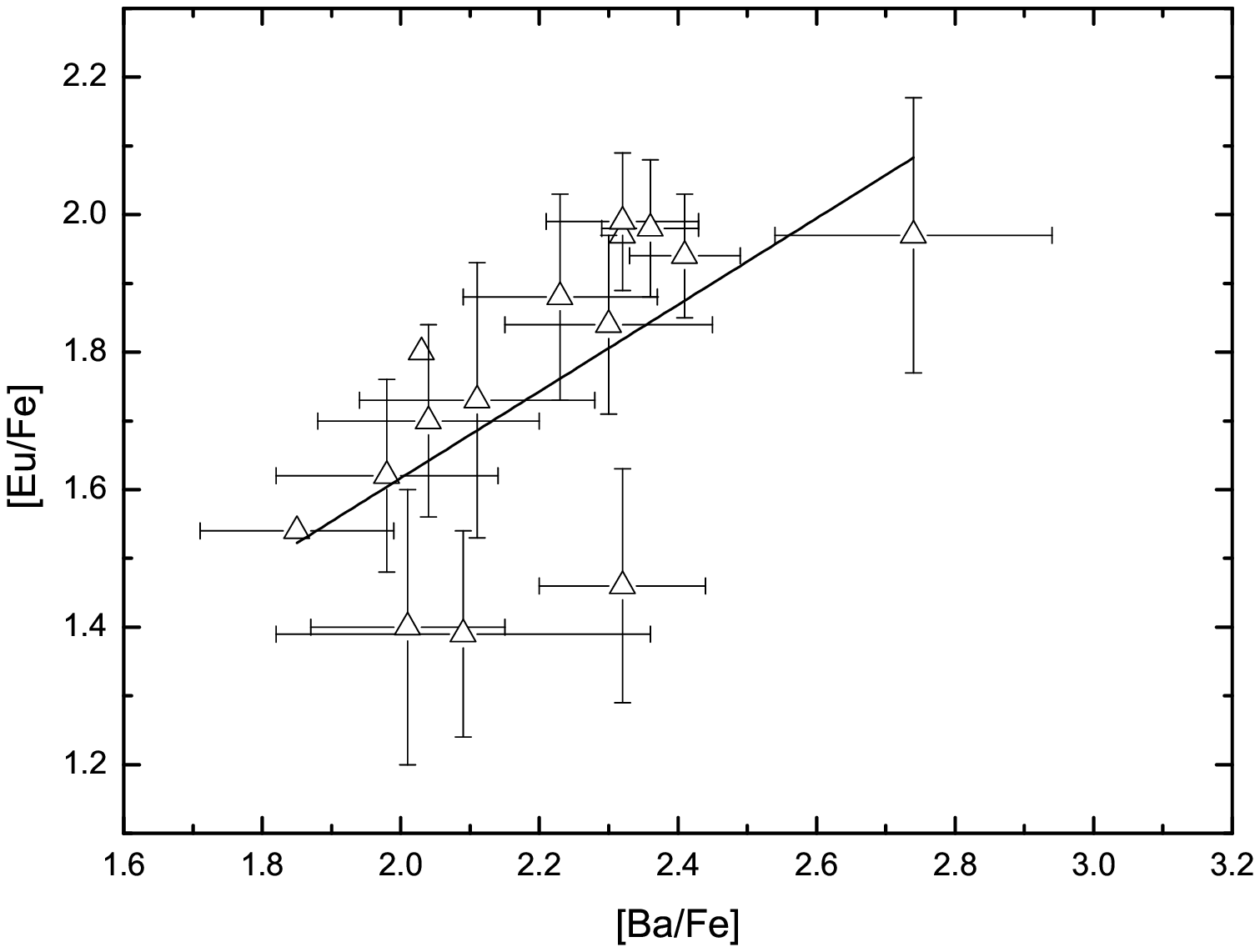}
 \hspace{-4.0cm}
 \includegraphics[width=.60\textwidth,height=.40\textheight]{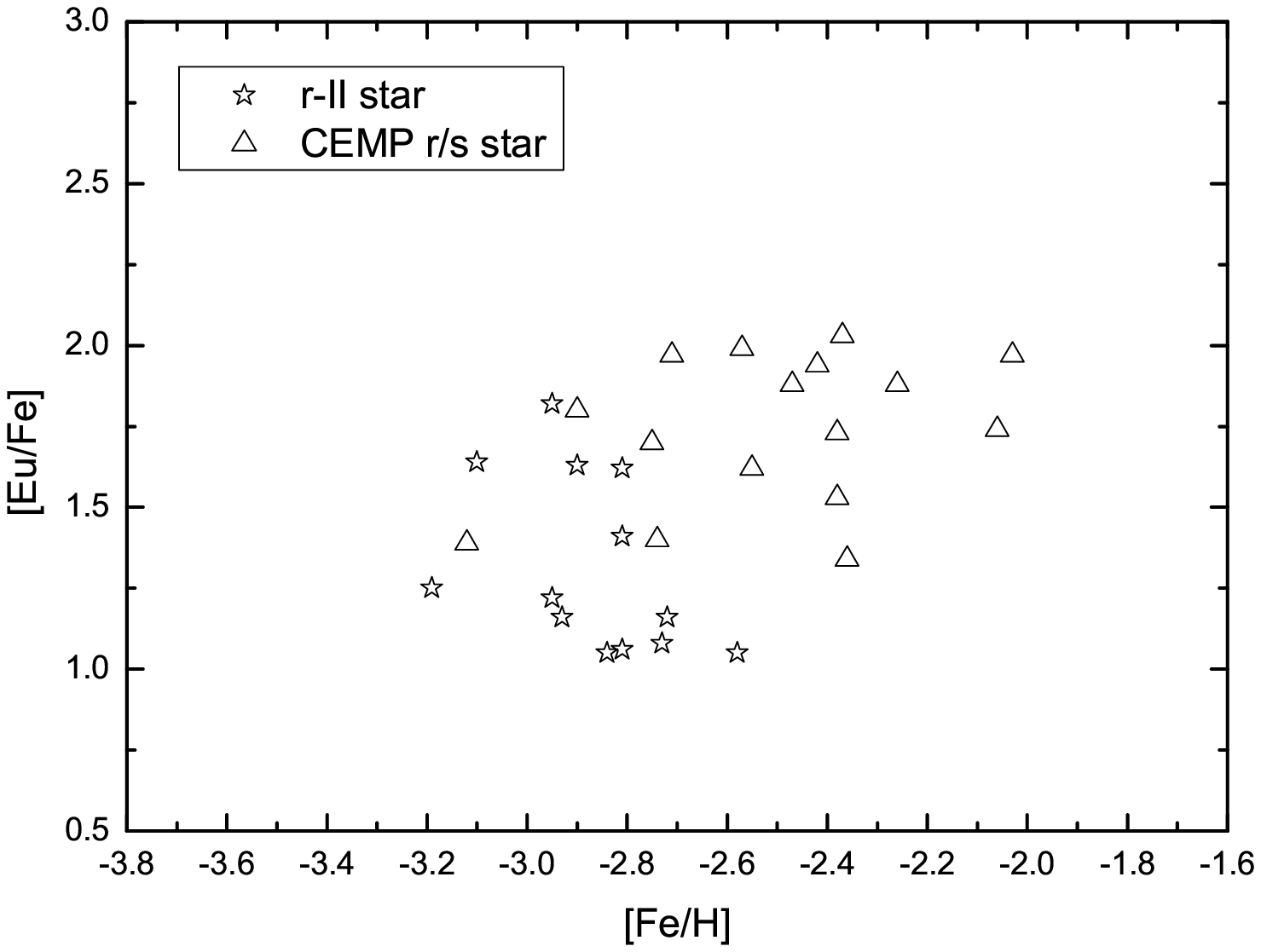}
  \vspace{-3.2cm}
 \begin{center}
  \caption{(a) [Eu/Fe] vs. [Ba/Fe] for CEMP-r/s stars; (b) [Eu/Fe] vs. [Fe/H].
  Where the unfilled triangles and stars represent CEMP-r/s and r-II stars, respectively. \label{Fig:euba}}
  \end{center}
\end{figure*}

Figure~\ref{Fig:euba}a shows the abundance ratio [Eu/Fe] versus
[Ba/Fe] for CEMP-r/s stars. An obvious correlation between [Eu/Fe]
and [Ba/Fe] can be seen from this figure.
It is strange that, as we know, Eu and Ba were produced by two
independent process (i.e. r- and s-process), respectively. For the
formation of CEMP-r/s stars, a popular mechanism is the r-elements
pre-enrichment, i.e. r-process material firstly polluted the
molecular cloud before the formation of such type stars. It should
be noted that the initial r-enrichment does not affect the s-process
nucleosynthesis \citep{bis11}. To explain this correlation, we have
to expect a similar correlation between the dilution event for r-
and s-elements during the mass transfer. \citet{all12}, however,
thought the more reasonable view is that only one process, i.e. the
s-process, is responsible for the production of both Ba and Eu,
because the AGB stars can produce quite high [Eu/Fe]. In fact, there
is a difficulty that whether the s-process could be able to produce
a reasonable proportion among the 2nd and 3rd r-peak elements such
as Te, Eu, Os, Pt and so on for CEMP-r/s stars \citep{bis12}.

We collected here almost all (about 13) r-II stars ([Eu/Fe]$>$1.0
and [Ba/Eu]$<$0, see \citealt{bee05}) observed up to now: CS
31082-001 \citep{ple04}, CS 22892-052 \citep{sne03}, CS 22183-031
\citep{hod04}, CS 22953-003 \citep{fra07}, CS 29491-069, HE
1219-0312 \citep{hay09}, CS 29497-004, HE 0430-4901, HE 0432-0923,
HE 1127-1143, HE 2224+0143, HE 2327-5642 \citep{bar05}, HE 1523-0901
\citep{fre07}. For comparison, we plot in figure~\ref{Fig:euba}b
[Eu/Fe] vs. [Fe/H] of these r-II stars and CEMP-r/s stars. From
figure~\ref{Fig:euba}b, we can see different dependence of [Eu/Fe]
on metallicity between CEMP-r/s and r-II stars. [Eu/Fe] of r-II
stars decreases with [Fe/H], and reaches a small [Eu/Fe] value
(about 1.05) at [Fe/H]\,$\approx-$2.5. The up [Eu/Fe] limit of
CEMP-r/s stars, however, still keep a high value (about 1.97)
unchanged up to [Fe/H]\,$\approx-$2.0. Furthermore, [Eu/Fe] of r-II
stars span a smaller range of [Fe/H], i.e. $-3.2<$\,[Fe/H]\,$<-2.5$,
than that of CEMP-r/s stars which is from [Fe/H]\,$=-3.1$ to $-2.0$.
This means that the r-process responsible for r-II and CEMP-r/s
stars should have different physical conditions. The SN\,II
accompanied with r-process maybe have different yields and initial
mass. The r-process yields responsible for CEMP-r/s stars should
increase with [Fe/H] in order to keep [Eu/Fe] unchanged. For r-II
stars, however, the r-process yields of SNII maybe remain stable,
the decreasing should be due to the increasing iron abundance in the
interstellar medium (ISM hereafter). The different [Fe/H] range of
r-II and CEMP-r/s stars maybe is the reason of the small number of
r-II stars observed up to now.

\section{The Parametric Model}
Large samples of s-enhanced stars with wide metallicity ranges,
including barium stars, CEMP-s stars, CEMP-r/s stars, can provide
a whole information about the nature of s-process. Abundance
signatures of heavy elements can help to identify the environment
of s-process nucleosynthesis regions \citep{cow06,sne08}. As the
 physical mechanism of the proton mixing
from the hydrogen-rich envelope to the $^{12}$C-rich layer to form
a $^{13}$C pocket \citep{bus01} is still unknown, the parametric
studies are still very useful to explain the formation of
s-enhanced stars. In this paper, we adopted an parametric model developed by
\citep{zha06,cui10} to simulate the abundance patterns of barium,
CEMP-s and CEMP-r/s stars in order to provide further insight into
the nature of s-process nucleosynthesis in the Galaxy.

Four parameters are needed by the parametric model for calculating
the s-process nucleosynthesis. They are the neutron exposure per
thermal pulse $\Delta\tau$, the overlap factor $r$, the component
coefficient of the s-process $C_{s}$ and the component coefficient
of the r-process $C_{r}$, where
$\Delta\tau=n_{n}\upsilon_{T}\Delta t$, and $\upsilon_{T}$ is the
average thermal velocity of neutrons at $10^8$ K (the appropriate
temperature for the $^{13}$C$(\alpha,n)^{16}$O reaction working).
In this model, the abundance of the $i$th element in the observed
star is calculated as follows:
\begin{equation}
\centering
N_{i}(Z)=C_{s}N_{i,\ s}+C_rN_{i,\ r}10^{[Fe/H]} ,
\end{equation}
where $Z$ is the metallicity of the star, $N_{i,\ s}$ is the
abundance of the $i$th element produced by s-process and $N_{i,\
r}$ is the abundance from r-process, $C_s$ and $C_r$ are the
component coefficients that correspond to contributions from the
s-process and the r-process, respectively (see Zhang, Ma \& Zhou,
2006 for details).

Barium stars belong to the stellar population I, which were formed
in a similar condition with the solar system. In other words,
these stars were `rich' in metallicity as soon as they were
formed. Thus, we adopted the scaled solar element abundances based
on [Fe/H] of barium stars as the initial abundances of seed
nuclei, and the same method were adopted for the initial seed
nuclei lighter than the iron peak elements of the CEMP-s and
CEMP-r/s stars. In the metal-poor environment, however, the
neutron-capture elements in molecular clouds, where CEMP-s and
CEMP-r/s stars formed, were mainly produced by r-process.
Generally, the r-process abundance pattern of stable
neutron-capture elements is divided into two parts, i.e. the
r-elements with $Z<56$ and $Z\geq56$. The r-elements with
$Z\geq56$ are mainly produced by the main-r process, while the
lighter ones are mainly contributed by a light element primary
process (LEPP) or/and weak-r process
\citep{tra04,kra07,mon07,cow11}. The main-r process produced a
stable abundance pattern from star-to-star for elements with
$56\leq Z<83$, which is consistent with a scaled solar r-process
pattern \citep{mon07}. Thus, the scaled
solar r-process abundance \citep{arl99} were adopted as
the initial abundance of the heavy
elements ($Z\geq56$). To the light nuclei
($Z<56$), it is still not clear that whether the
r-process pattern is stable from star-to-star, because of the
still limited amount of stellar observations for these elements up
to now \citep{cow11}. Thus, for the light nuclei we adopted the
method suggested by \citet{zha06}.

During the calculations, the chi-square (i.e. $\chi^{2}$) test was
also applied in order to get the best simulation results, and the
parameters of s-process nucleosynthesis region.

\section{Results and Discussion}

\subsection{Abundance Analysis for Barium Stars}
Using the observed abundance data of 35 barium sample stars
\citep{all06a,smi07,pom08}, the model parameters can be obtained.
The resulted parameters for the neutron exposures per pulse, the
overlap factor, the mean neutron exposures and the component
coefficients for r- and s-process are listed in Tables 1 and 2.

\begin{figure*}
 \centering
 \includegraphics[width=1.6\textwidth,height=1.5\textheight]{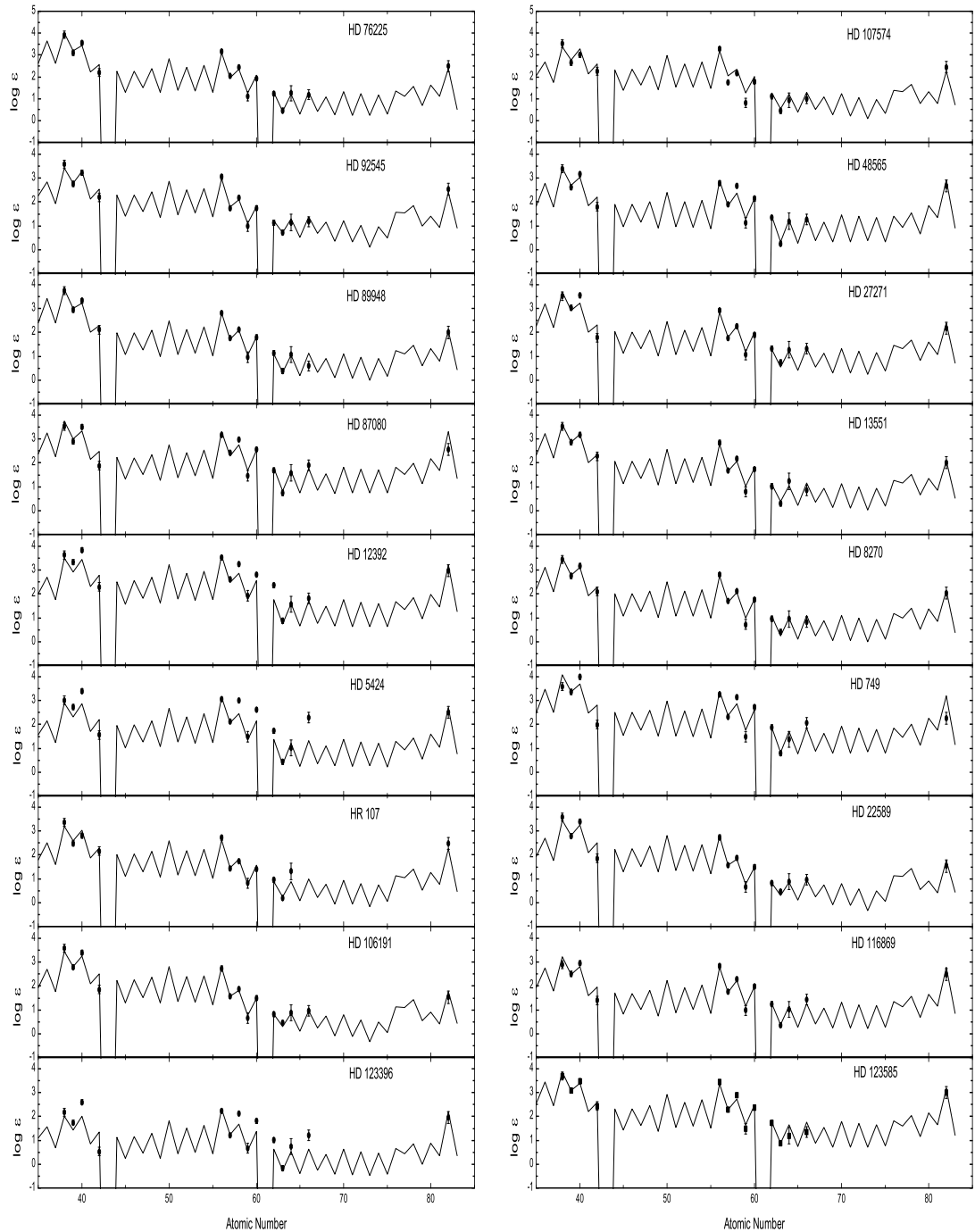}
 \vspace{-14.5cm}
 \caption{\label{Fig:bafit} Best fits to the observational results of barium stars. The filled black
 circles with appropriate error bars denote the observed element abundances;
 the solid lines represent predictions from s-process calculations,
 where the r-process contribution is considered simultaneously.
 Spectroscopic data adopted from \citet{all06a,smi07,pom08}.}
\end{figure*}

The best simulation results (solid lines) for each barium star are
shown in figure~\ref{Fig:bafit}, and the observation data (black filled circles)
are also plotted for comparing. In figure~\ref{Fig:bafit}, 26 barium stars are
presented as log $\varepsilon$(X)\footnotemark[1]
\footnotetext[1]{log $\varepsilon$(X) = log$(N_{X}/N_{H})+12$.}
versus atomic numbers, while the rest nine barium stars are shown
as [X/Fe] versus atomic numbers. We can see that the curves
calculated by the parametric model, to most stars, agree with the
observed abundances well within the error bars. The good agreement
of the model results with the observations supports the parametric
model we used.

The overlap factor, $r$, denotes the mass fraction of part
material in the s-process nucleosynthesis region, which survived
the third dredge-up event and then underwent the succeeding
neutron exposure. The $r$ values of the AGB companions
of 35 barium stars belong to the range $0.02\sim0.39$. There are
some stars, such as HD 5424 and HD 107574, whose $r$ values are
lower than $0.1$, which means that few seed nuclei could
experience successive neutron exposure, in other word, the third
dredge-up event is very efficient in this case. This situation can
be regarded as the so-called single exposure event. \citet{all06b}
also obtained a single exposure for HD 5424 and HD 107574 as a
best fit to their s-patterns.

\begin{figure*}
\centering
 \includegraphics[width=0.8\textwidth,height=.4\textheight]{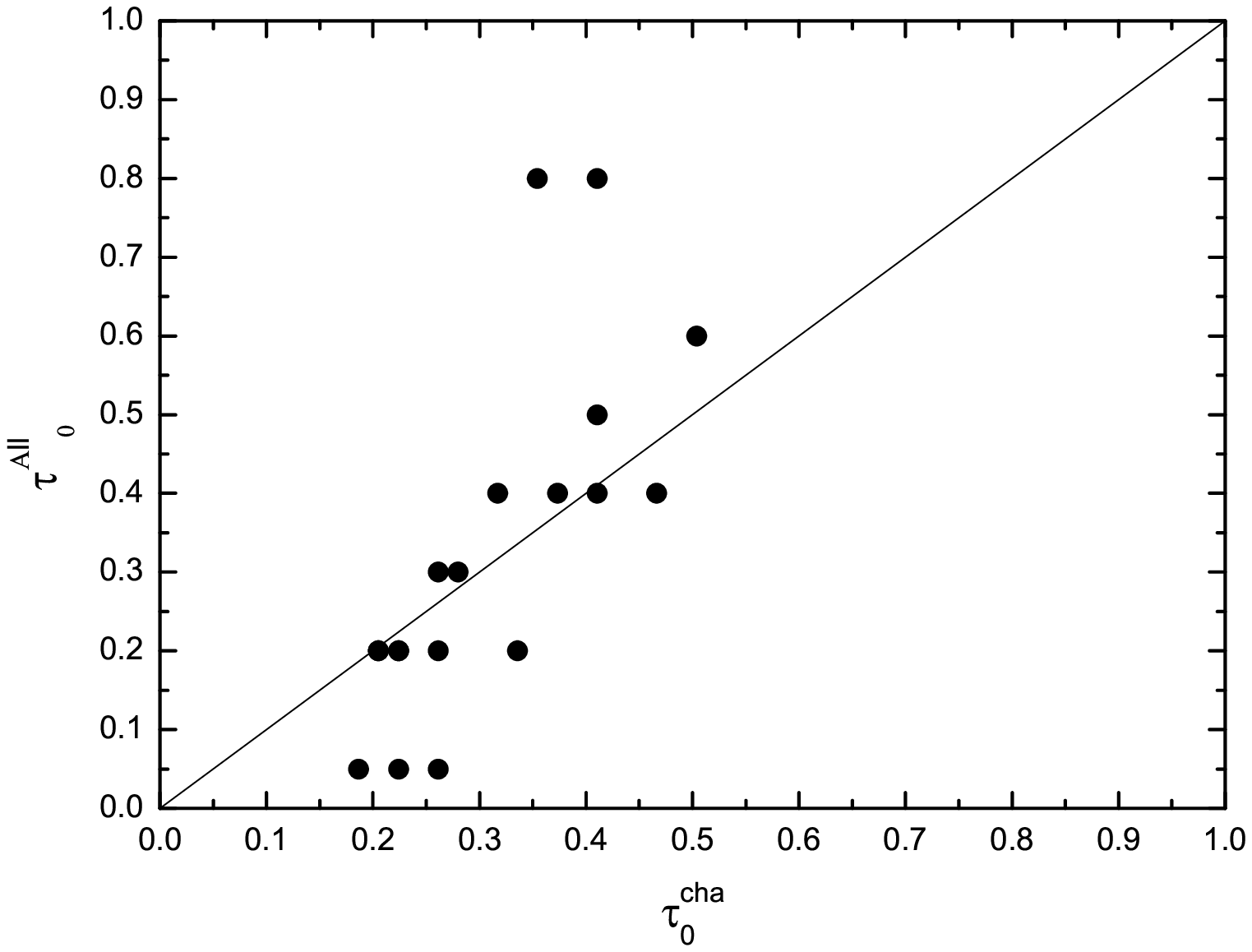}
 \suppressfloats[t]
  \renewcommand{\thefigure}{7}
 \vspace{-3.4cm}
 \caption{\label{Fig:cmne}Comparing the mean neutron exposure of barium stars with results of \citet{all06b}.}
\end{figure*}

In the case of multiple subsequent exposures, the mean neutron
exposure is given by $\tau_{0}=-\Delta\tau/$ln$r$. The mean
neutron exposures of the AGB stars, $\tau_{0}$, corresponding to
the 35 barium stars were also shown in tables 1 and 2. In order to
compare our results with those of \citet{all06b}, we convert our
$\tau_{0}$ values into $\tau^{cha}_{0}$ based on the the formula,
$\tau^{cha}_{0}=\tau_{0}\times(T_{9}/0.348)^{-1/2}$ (where
$T_{9}=0.1$, in units of 10$^{9}$ K). The results are plotted in
figure~\ref{Fig:cmne}, where $\tau^{All}_{0}$ was taken from \citet{all06b}.
Based on the definition of $\tau_{0}$, there is no a real mean
neutron exposure in the single exposure mechanism. Thus the mean
neutron exposure values relating to 19 barium stars were included
in figure~\ref{Fig:cmne} for comparing, while seven barium stars were excluded,
such as HD 5424, HD 107574, etc, where the single exposure event
was responsible for the s-process \citep[see also][]{all06b}.
General agreement can be found for our results to those of
\citet{all06b}, while there are five barium stars, the obvious
difference between $\tau^{cha}_{0}$ and $\tau^{All}_{0}$ can be
seen. Three of them, i.e. HD 27271, HD 106191, and HD BD+18 5215,
the small value 0.05 mbarn$^{-1}$ was calculated based on an
exponential exposure under the neutron density $10^{12}$ cm$^{-3}$
\citep{all06b}, this density is much higher than that of the
prediction of \citet{gal98} for the $^{13}$C neutron source, while
the other two barium stars, i.e. HD 749 and HD 12392, the value of
$\tau^{All}_{0}$ is 0.80 mbarn$^{-1}$, which is higher than our
results, but \citet{all06b} also give 0.406 mbarn$^{-1}$ and 0.40
mbarn$^{-1}$ as values of $\tau^{All}_{0}$($\sigma$N), those are
close to our results, i.e. 0.354 mbarn$^{-1}$ and 0.41
mbarn$^{-1}$, respectively.

\subsection{Abundance Analysis for metal-poor Stars}
As mentioned above, our sample include 24 metal-poor stars, i.e.
CEMP-s and CEMP-r/s stars here. Among them, 15 stars have been
studied in our previous works \citep{zha06,cui07a,cui07b,cui10},
while the other nine stars were simulated in this work using the
parametric model. The nucleosynthesis parameters for these nine
stars were shown in table 3, and for comparing reason the
parameters for the other 15 stars were also presented.

\begin{figure*}
 \centering
 \includegraphics[width=1.6\textwidth,height=1.2\textheight]{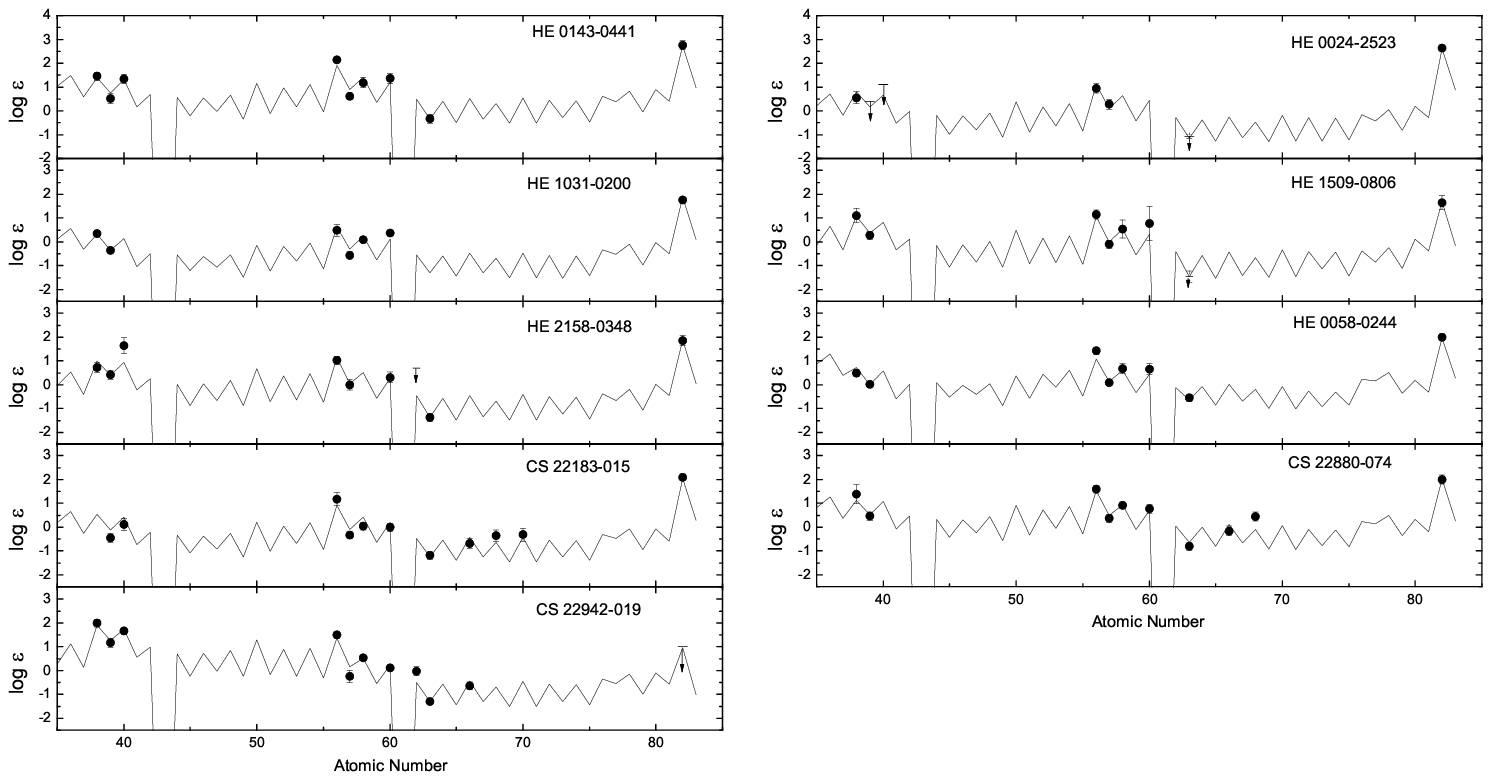}
  \suppressfloats[t]
 \renewcommand{\thefigure}{8}
 \vspace{-15.5cm}
 \caption{\label{Fig:mpfit}Best fits to the observational results of metal-poor stars.
  The symbols are same with figure~\ref{Fig:bafit}.}
\end{figure*}

Figure~\ref{Fig:mpfit} shows our best fits to the observational results of the
nine metal-poor stars. The black filled circles with error bars
denote the observed element abundances, while the solid lines
represent the predictions from s-process calculations, in which
the r-process contribution is considered simultaneously. The
spectroscopic data adopted from \citet{coh06,luc03,joh02,aok02}.
We can see that for most stars the calculated results fit with the
observed data well within the error bars. The good simulation
results suggest that our parametric model is also valid for
metal-poor stars.

The $r$ values for most of the metal-poor stars are in the range
$0.1\sim0.86$, while two stars, i.e. HE 1305-0007 and HD 189711,
have $r<0.10$. The lower $r$ value means that only a few seed
nuclei can receive the succeeding neutron exposure during the next
interpulse phase, in other words, the efficiency of the third
dredge-up in their former AGB companions is very high. This case
also belongs to the so-called single exposure event. We note this
situation is also happened for barium stars as discussed above,
thus, we can see that the single exposure event for the s-process
nucleosynthesis is general in a wide metallicity range of our
Galaxy. Combining the analytical formula of the overlap factor
given by \citet{ibe77} with the initial-final mass relation for
AGB stars \citep{zij04}, \citet{cui06} derived a function of $r$
varying with metallicity and initial mass of AGB stars. Using this
function, we can find the $r$ range of the AGB stars with initial mass range
$1.0\sim4.0$ M$_{\odot}$ is very similar with that found
in this work excluding the two stars, HE 1305-0007, and HD 189711.

The neutron exposure per pulse, $\Delta\tau$, is also a fundamental
parameter in the s-process nucleosynthesis. The calculated values
of $\Delta\tau$ vary from 0.23 mbarn$^{-1}$ to 0.88 mbarn$^{-1}$.
We compared the $\Delta\tau$ value with the one we can get
from the literature for two stars, LP 625-44 and LP 706-7. Our
values, 0.69 mbarn$^{-1}$ and 0.82 mbarn$^{-1}$ are very close to
their ones, 0.71 mbarn$^{-1}$ and 0.80 mbarn$^{-1}$ \citep{aok01},
respectively. The mean neutron exposure ($\tau_{0}$), here, is
even a more important parameter for the s-process nucleosynthesis
calculation, whose value mainly dominate the final abundance
distribution of the s-process elements in AGB stars. In general,
the larger value of $\tau_{0}$, the more heavier elements can be
produced. For example, the lead stars produced in the low
metallicity environments where the higher $\tau_{0}$ exists.

\subsection{s-Process Nucleosynthesis Characters at Different Metallicities}
In order to get a broader view of the s-process nucleosynthesis,
we compare the physical parameters relating to metal-poor stars
and barium stars together in the following text, especially to
identify the impact of metallicity.

\begin{figure*}
 \centering
 \includegraphics[width=.60\textwidth,height=.40\textheight]{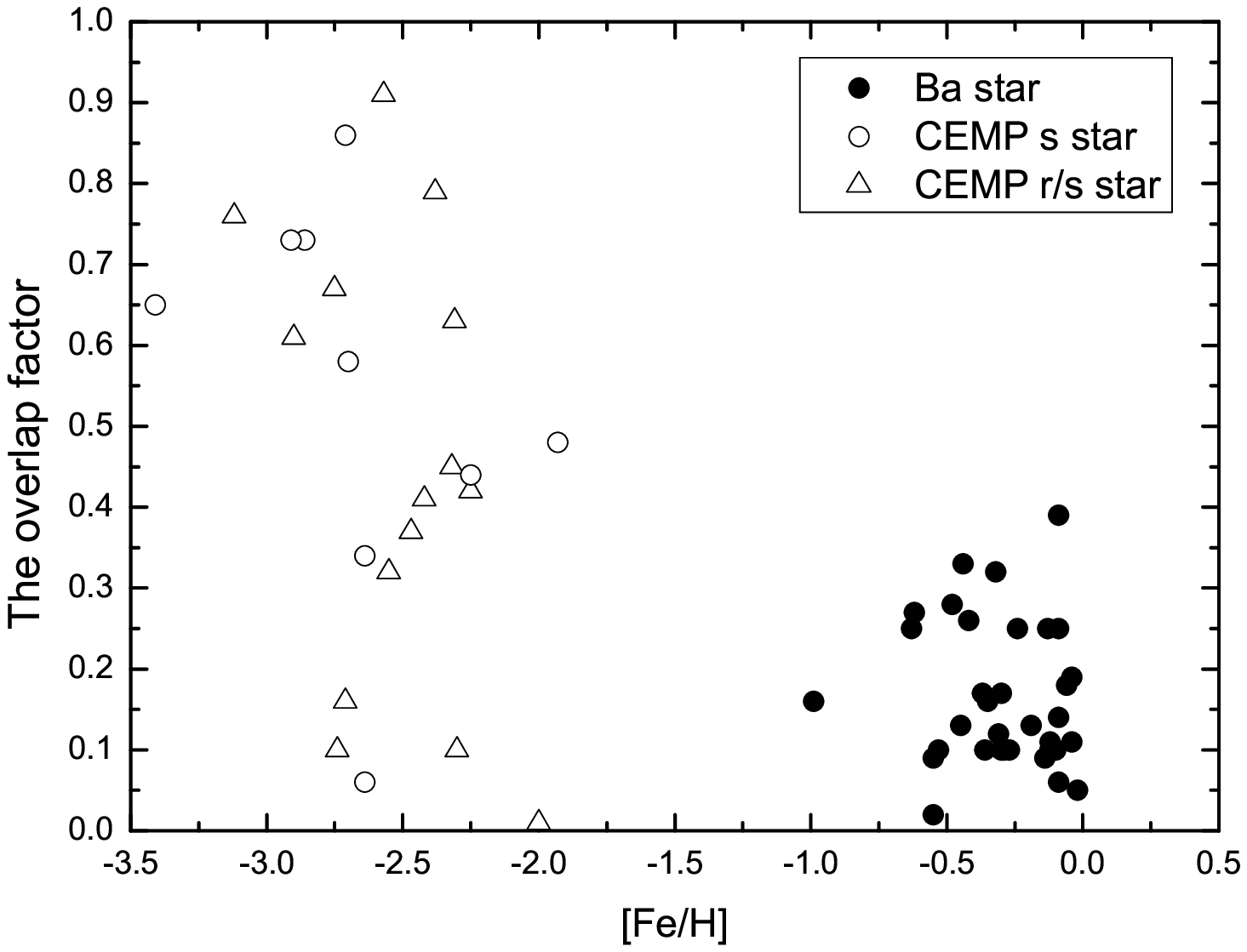}
 \hspace{-4.0cm}
 \includegraphics[width=.60\textwidth,height=.40\textheight]{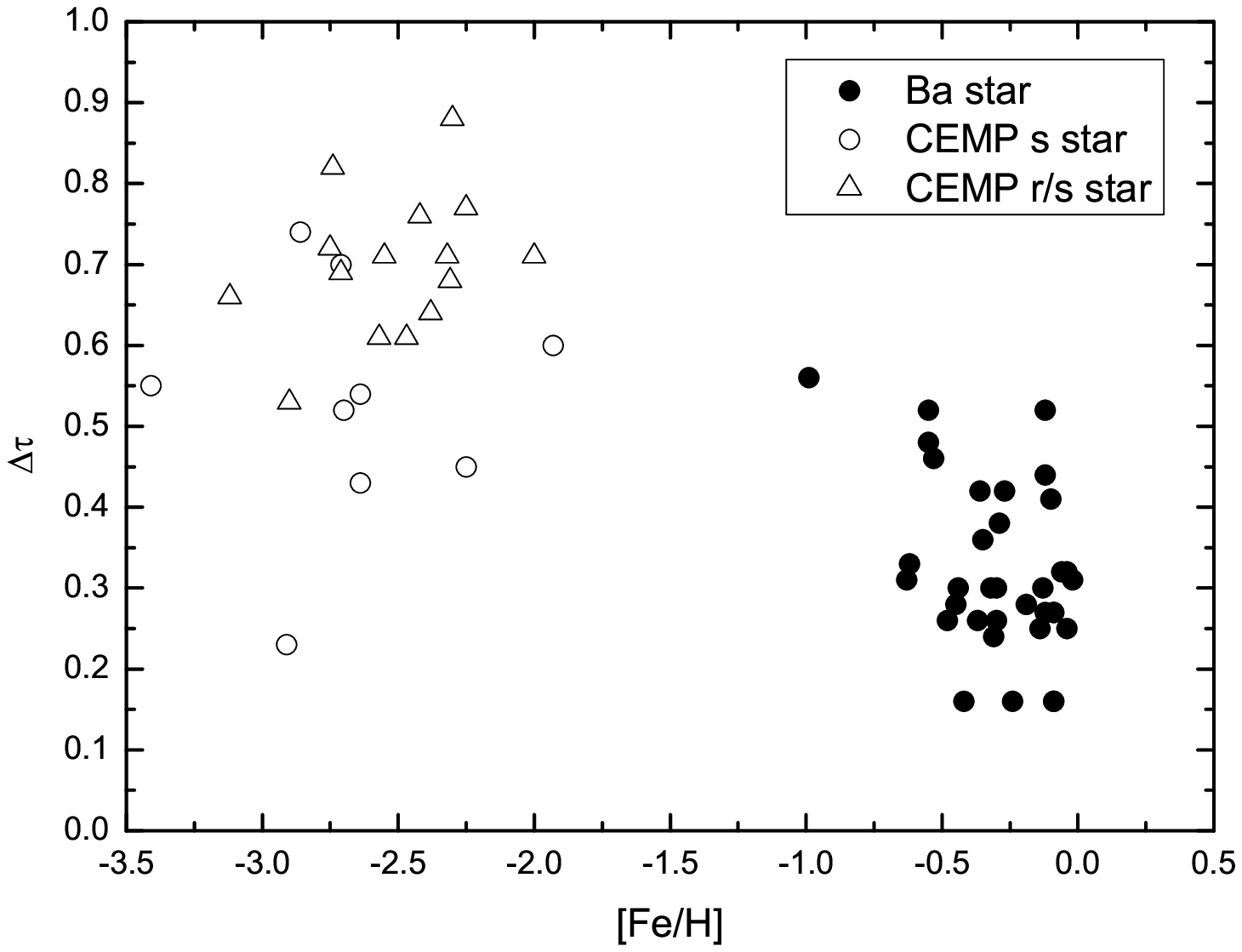}
  \renewcommand{\thefigure}{9}
 \vspace{-3.2cm}
 \caption{\label{Fig:ofnepp}(a) the overlap factor, $r$, versus [Fe/H]; (b) the neutron exposure
 per pulse, $\Delta\tau$, versus [Fe/H]. The meaning of symbols are same in figure~\ref{Fig:XFe}.}
\end{figure*}

Figure~\ref{Fig:ofnepp} shows $r$ and $\Delta\tau$ versus [Fe/H], respectively.
The filled circles represent barium stars, and the unfilled
circles represent CEMP-s stars, while the unfilled triangles
represent CEMP-r/s stars. We can see that the range of $r$ for
metal-poor stars ($0.01\sim0.86$), is larger than
that for barium stars ($0.02\sim0.39$). Obviously, the s-process
nucleosynthesis is more complicated in metal-poor stars. It should
be noted that most $r$ values of our metal-poor samples are larger
than 0.40 (see figure~\ref{Fig:ofnepp}a). As discussed above, the larger the
overlap factor value is, the more iron seeds can receive repeated
neutron exposure, which is in favour of heavier elements
producing, such as Ba, Pb, etc. \citet{gal98} gave a $r$ range,
i.e. $0.40<r<0.70$, using their dedicated evolutionary model for
AGB stars with initial mass from 1 to 3 $M_{\odot}$ and
metallicity from solar to half solar. Obviously, they are larger
than ours for barium stars. Nevertheless, it should be kept in
mind that, in Gallino et al.'s model, only $0.05$ fraction
material can really receive the neutron exposure each time.

From figure~\ref{Fig:ofnepp}b, we can see that $\Delta\tau$ value increases with
decreasing metallicity, which also support the results of favoring
heavier elements production, due to the higher ratio of neutron to
iron seeds in low metallicity environment. \citet{mat92} presented
a formula, i.e. $n_{n}\propto1/(a+Z/Z_{\odot})$, to describe the
equilibrium neutron density for the primary neutron source
$^{13}$C($\alpha$, n)$^{16}$O, where the constant $a$ is roughly
given by the ratio of the average neutron-capture cross section
times abundance for newly synthesized elements (i.e. so-called
primary neutron poisons, such as $^{16}$O, $^{14}$N, etc.) to the
average capture cross section times solar abundance for heavy
initial elements, usually $a\sim0.001$ is adopted. \citet{bus99}
and \citet{gal99} provided a more simplified relationship between
the neutron density and metallicities, i.e. the typical neutron
density in the nucleosynthesis zone scales roughly as $1/Z^{0.6}$,
for $Z_{\odot} \geq  Z \geq 0.02Z_{\odot}$. At low metallicities,
the effect of the primary poisons prevails. The tendency of
neutron density varying with metallicity can be seen clearly from
figure~\ref{Fig:ofnepp}b. Of course, the neutron exposure time must be considered
simultaneously. In addition,   no apparent distribution difference
between  $r$ and $\Delta\tau$ values can be found for both CEMP-s
and CEMP-r/s stars.

\begin{figure*}
 \vspace{2.2cm}
 \centering
 \includegraphics[width=.60\textwidth,height=.40\textheight]{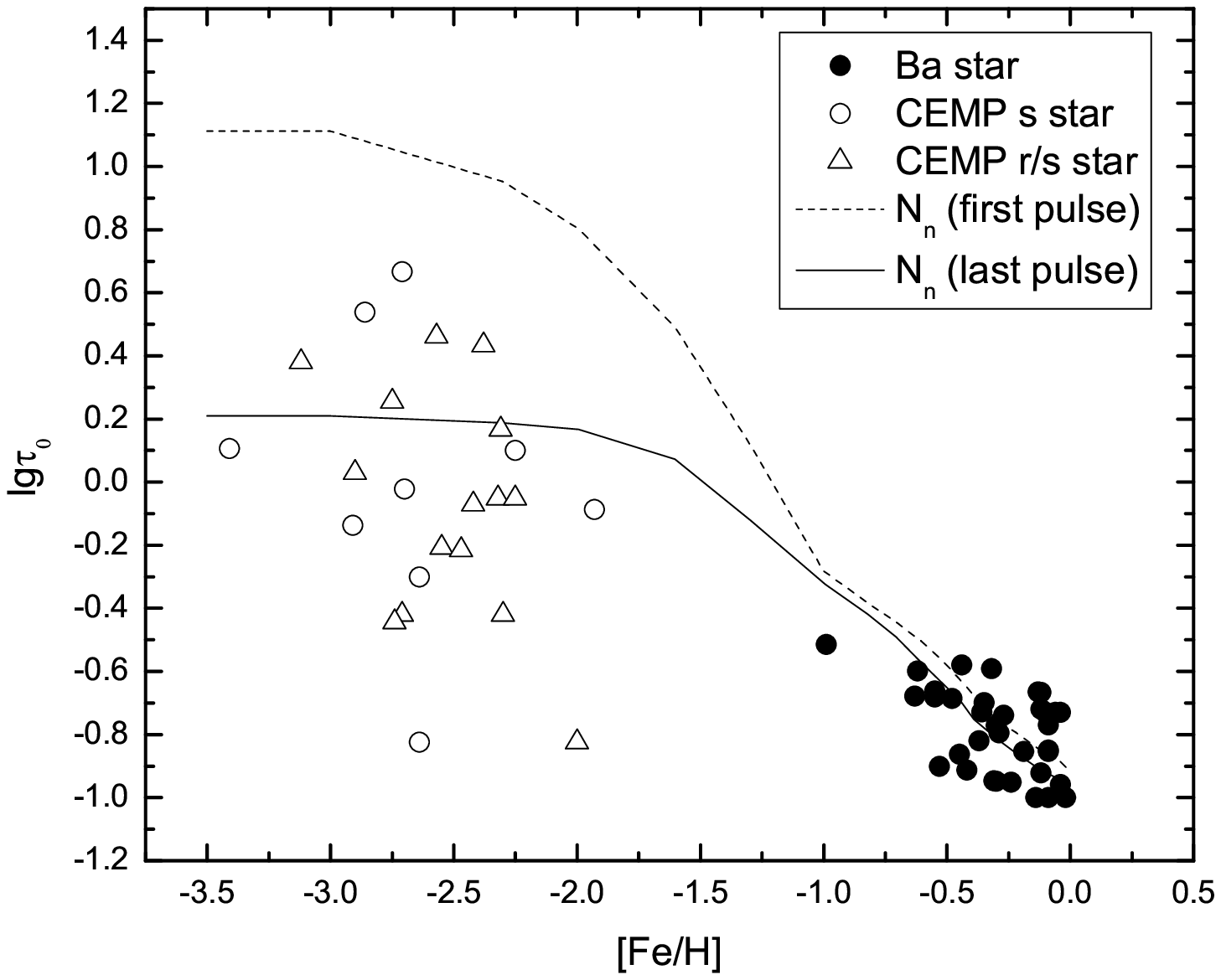}
  \renewcommand{\thefigure}{10}
 \vspace{-3.2cm}
 \caption{\label{Fig:mneFeH}lg$\tau_0$ versus [Fe/H]. The meaning of
 symbols are same in figure~\ref{Fig:XFe}, except the dotted line for neutron
 density (lg$N_{n}$) after the first TDU,
 and solid line for the other subsequent neutron density.}
\end{figure*}

As we know, $\tau_{0}$ represents the synthetical effect of $r$
and $\Delta\tau$, and determines the final abundance distribution
of heavy elements in AGB stars, thus, it's a more important
parameter here. We present lg$\tau_0$ versus [Fe/H] in figure~\ref{Fig:mneFeH},
and a larger scatter exists in lg$\tau_0$ for CEMP-s and CEMP-r/s
stars than that in barium stars can be seen, and the lg$\tau_0$
values of CEMP-s and CEMP-r/s stars are generally larger than
those of barium stars. Obviously, the efficiency of s-process
nucleosynthensis responsible for heavy neutron-capture elements of
metal-poor samples is higher than that of barium stars. In other
words, the more enhanced heavy elements, especially for Pb, can be
reached at the lower metallicity environments, which has be proved
by the observation data shown in figure~\ref{Fig:XFe} and ~\ref{Fig:Pbhs},
especially, the [Pb/hs] values of metal-poor stars are almost larger than 0,
while the barium stars on the contrary.
Of course, this could not work at a ultra low-metallicity
environment, because the medium- and low-mass stars hadn't evolved to AGB phase,
which is thought as the most appropriate site for s-process nucleosynthesis.

From figures~\ref{Fig:ofnepp} and ~\ref{Fig:mneFeH} we can see that the three important
parameters refer to s-process nucleosynthesis, i.e. $r$,
$\Delta\tau$, $\tau_0$, have similar distribution range in both CEMP -s
and CEMP-r/s stars. This supports a uniform origin of the
s-process elements for both CEMP-s and CEMP-r/s stars. Moreover,
as the $\tau_0$ dominated the final element abundance
distribution, the AGB companions should have similar s-process
efficiency when $\tau_0$ is analogous. If the r-process
contribution were taken away from CEMP-r/s stars, a similar dispersion of
abundance ratios, such as [hs/Fe], [Pb/Fe] and [Pb/hs], with
CEMP-s stars should be expected. Obviously,
the r-process should be responsible for the abundance excess of
Ba, Pb, etc. This supports the conclusions derived from the similar
proportion of r-process contribution for solar Ba and Pb in
section 3. This means that we maybe not need a third process
to explain the features of both r- and s-process for CEMP-r/s stars \citep{lug12}.

For more information, the curves of neutron density (i.e.
lg$N_{n}$) varying with metallicities given by \citet{gal99} were
also plotted in figure~\ref{Fig:mneFeH}. Where the dotted line represents the
maximum neutron density during the interpulse phase after the
first occurrence of TDU, and the solid line for a late interpulse
period. The curves have been normalized to the neutron density
corresponding to $\tau_{0}=0.152$ mbarn$^{-1}$, which can give the
main s-process component of the solar system using a AGB model with half solar metallicity
(i.e. [Fe/H]$\simeq-0.3$). Although there is a large scatter, the
tendecy of lg$\tau_{0}$ increasing with decreasing metallicities
can be seen, which implies that the neutron density plays a more
important role in the s-process nucleosynthesis of AGB stars.

\begin{figure*}
 \centering
 \includegraphics[width=.60\textwidth,height=.40\textheight]{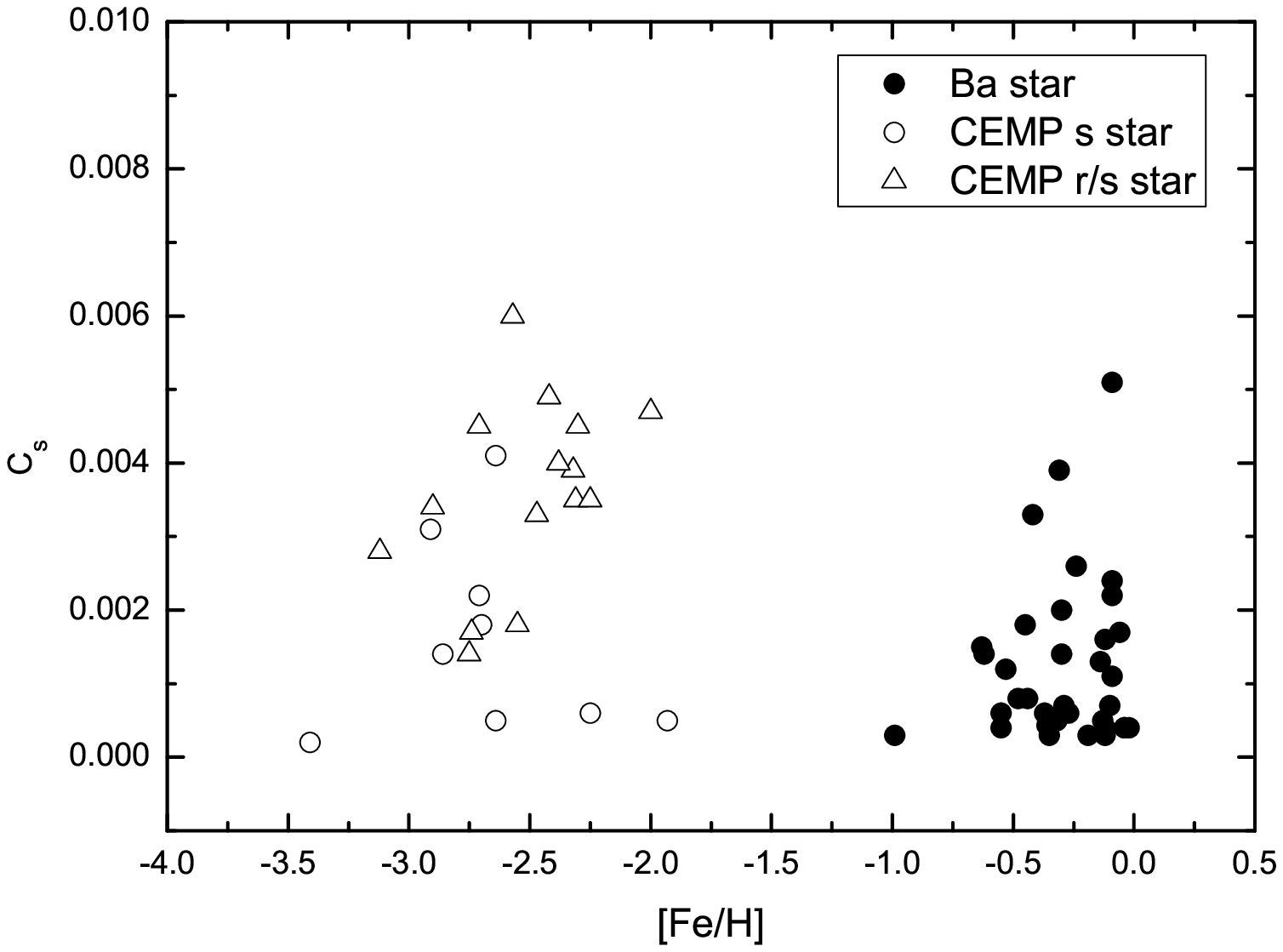}
 \hspace{-4.0cm}
 \includegraphics[width=.60\textwidth,height=.40\textheight]{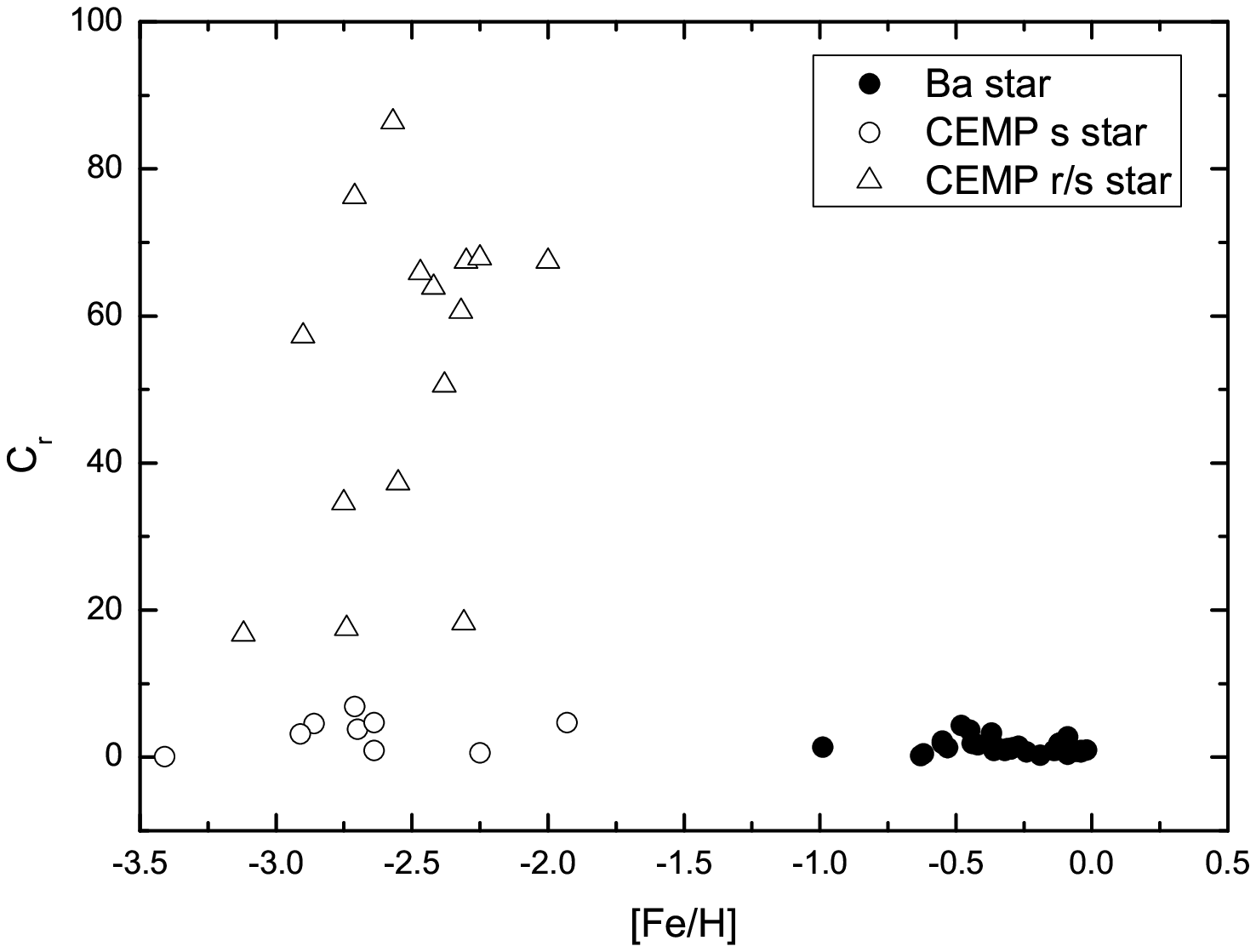}
  \renewcommand{\thefigure}{11}
 \vspace{-3.2cm}
 \caption{(a) the component coefficient of the s-process, $C_{s}$, versus [Fe/H]; (b) the component coefficient
of the r-process, $C_{r}$, versus [Fe/H]. The meaning of symbols are same in figure~\ref{Fig:XFe}.\label{Fig:CFeH}}
\end{figure*}

The component coefficient of the s-process, $C_{s}$, and the
component coefficient of the r-process, $C_{r}$, versus [Fe/H],
are presented in figure~\ref{Fig:CFeH}a,b, respectively. The distribution range
of $C_{s}$ for metal-poor stars is close to that of barium stars
(Figure~\ref{Fig:CFeH}a). It's well known that, $C_{s}$ is related to some
important evolution characters of AGB stars, such as structure,
efficiency of the TDU, mass of convective envelope, etc. In addition, $C_{s}$
is also related to the physical parameters of binary system, which
the CEMP-s, CEMP-r/s stars or barium stars and their pre-AGB
companions belong to, such as the orbital periods, mass accretion
efficiency and so on \citep[see the definition of $C_{s}$
by][]{zha06,cui07b}. Thus, it can be explained that the total
dilution efficiency of s-enhanced material from s-process
nucleosynthesis region to the star observed now, should be very
similar between metal-poor stars and barium stars. In other words,
the influence of metallicity is not important on the total
dilution efficiency of s-element abundances, which also supports
that the s-element enhancement should be resulted by mass transfer
from AGB stars for CEMP-s, CEMP-r/s and barium stars. For most of the barium
stars, the binarity have been confirmed \citep{mcc84,jef96,udr98a,udr98b}.

On the contrary, the $C_{r}$ values of CEMP-r/s stars are larger
than those of CEMP-s and barium stars, because the latter two kinds
are s-only stars. The small $C_{r}$ distribution ranges about
$0-6.9$ for CEMP-s stars and $0-4.3$ for barium stars imply that
their r-elements all come from the ISM, where the local polluting
event from SN II with r-materials didn't take place recently. While
the slightly larger scatter in CEMP-s stars supports the conclusion
that the ISM was not mixed well at the early epoch of our Galaxy
\citep{rya91}. The large scatter of $C_{r}$ for CEMP-r/s stars is
consistent with the observed large abundance dispersion of the
r-elements, such as Eu, Pt etc. However, it is not appropriate that
simply regarding CEMP-r/s stars as formed out of r-enriched clouds
because of their high frequency in metal-poor stars
\citep[e.g.][]{aok02}. \citet{qia01} also argued that the ISM was
sufficiently inhomogeneous to contain such great r-element
overabundance relative to Fe at [Fe/H] $>-3$. This scenario would
still be possible, however, if it is modified as that the formation
of the binary system is triggered by only one or a few supernova
which also provides the r-elements \citep{gal05,iva05,kap11,bis11}.
Then, the abundance dispersion could be a natural result from the
different r-elements yields of supernova progenitors with different
masses. The opinion that the exploding frequency of core-collapse
supernovae in early Galaxy is higher enough to explain the
observation results of old halo stars, is supported by the chemical
evolution model of our Galaxy \citep{ish99,ish04}. Although, the
chemical inhomogeneities of ISM locally polluted by short-lived
massive stars exploding as SN II is often used to explain the
r-elements origin of CEMP-r/s stars \citep{ots00,tho01,bis09,cui10},
many arguments on how core-collape supernovae produce r-process
materials still exists \citep{woo94,tho01,tho03,wan02}, the
AIC-pollution, 1.5 SN-pollution, etc are also possible origins of
the r-elements of CEMP-r/s stars (detailed discussion see Jonsell et
al. 2006 and references therein).

\begin{figure*}
 \centering
 \includegraphics[width=.60\textwidth,height=.40\textheight]{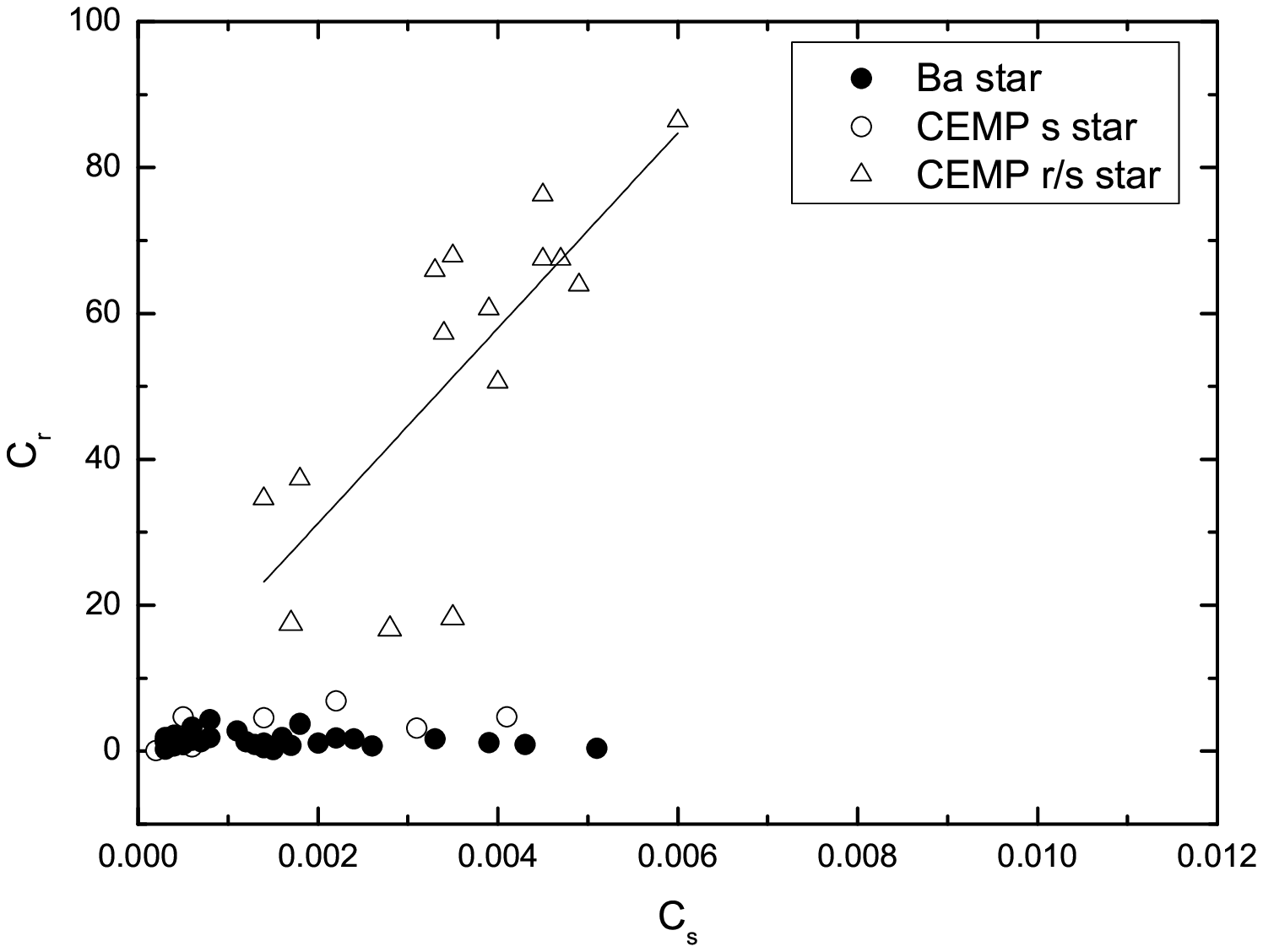}
 \suppressfloats[t]
  \renewcommand{\thefigure}{12}
 \vspace{-3.2cm}
 \caption{The relation between $C_{s}$ and $C_{r}$ of CEMP-s, CEMP-r/s and barium stars.
 The meaning of symbols are same in figure~\ref{Fig:XFe}.\label{Fig:crcs}}
\end{figure*}

In figure~\ref{Fig:crcs}, we plot $C_{r}$ versus $C_{s}$. Obviously, there is
no any correlation between $C_{s}$ and $C_{r}$ for s-only stars,
i.e. CEMP-s and barium stars, while, an obvious positive
correlation between $C_{s}$ and $C_{r}$ for CEMP-r/s stars as
presented in \citet{zha06} exists when more samples included.
\citet{zha06} suggested that this imply an increase of s-process
matter accreted from the AGB star with increasing r-process matter
accreted from the AIC or SN 1.5, which is a significant evidence
for the formation scenario of CEMP-r/s stars. Thus, it seems that
the AIC and SN 1.5 mechanisms still can not be dismissed,
especially for the CEMP-r/s stars with higher $C_{s}$ value.
In fact, the distribution of some fundamental parameters of binary
system, such as orbit radius, orbit period, and so on, should be
similar in the environments with different metallicities, and the
evolution features of AGB stars, such as structure, efficiency of
the TDU, mass of convective envelope, and so on, should also be similar for
CEMP-s and CEMP-r/s stars with similar metallicity. Thus,
the more probable scenario is that many origins of r-elements
exist together, i.e. the modified pre-enrichment is a common
origin for CEMP-r/s stars, which is sustained by \citet{bis11},
and AIC or SN 1.5 is supplementary, especially, for whose pre-AGB
companion with higher mass and small orbit radius which support
the higher $C_{r}$ and $C_{s}$ values, respectively.

Our model is based on the observed abundances of the s-enriched stars
and nucleosynthesis calculations, so the uncertainties of those observations
and measurement of the neutron-capture cross sections have been involved in the model calculations.

\section{Conclusions}
We have compared the abundance ratios of heavy elements of 35 barium
stars and 24 CEMP-stars including nine CEMP-s and 15 CEMP-r/s stars.
The tendency of [Pb/Fe] and [hs/Fe] increasing with decreasing
metallicity implies that the environments with lower metallicities
are favouring the heavier elements producing, especially for Pb
\citep{gal98,gor00}. And the larger scatter of the efficiency of
s-process nucleosynthesis in CEMP-s and CEMP-r/s stars than that in
barium stars was found because of lager scatter of their [Pb/hs]
ratios, while the similar distribution of [Pb/hs] between CEMP-s and
CEMP-r/s stars indicates that the s-process material of both CEMP-s
and CEMP-r/s stars should have a uniform origin, i.e. mass transfer
from their pre-dominant AGB companions in the corresponding binary
systems. Furthermore, for the CEMP-r/s stars, we found that the
r-process should provide similar proportional contributions to the
second s-peak and the third s-peak elements but marginal
contribution to the first s-peak elements if the the s-process
contribution were taken away, which should be responsible for the
higher over-abundances of heavy elements in CEMP-r/s stars than
those in CEMP-s stars. This hints that r-elements origin of CEMP-r/s
stars should be closely linked to the main r-process, especially for
the elements with $Z\geq56$.

Based on simulating the abundance distribution of heavy elements
of our sample stars, the physical parameters of their
corresponding s-process nucleosynthesis were derived. It is found
that there are almost larger values of $r$, $\Delta\tau$ and
$\tau_{0}$ for CEMP-s and CEMP-r/s stars than those for barium
stars, which supports the theory that there is higher efficiency
for heavier elements producing, especially for Pb, at lower
metallicities \citep{gal98,gor00}. In addition, the fact that some small $r$
values exist for both barium and CEMP-stars, implies that the
single exposure event of the s-process nucleosynthesis should be
general in our Galaxy. And the similar $C_{s}$ values between
metal-poor and barium stars implies that the total dilution
efficiency of s-enhanced material from s-process nucleosynthesis
region to the star observed now, should be very similar for both
type stars, i.e. the influence of metallicity is not important on
the total dilution efficiency of s-elements abundances. While, the
tiny difference of the $C_{r}$ values between CEMP-s and barium
stars, could be explained by the result that all the r-elements
come from the ISM, where the local polluting
event from SN II with r-materials didn't take place recently, and
the ISM was not mixed well at the early epoch of our Galaxy.
Combining the larger scatter of $C_{r}$ values and the positive
correlation between $C_{r}$ and $C_{s}$ for CEMP-r/s stars, we
suggest that the modified pre-enrichment scenario, i.e. the
formation of the binary system, which the CEMP-r/s star belongs
to, is triggered by only one or a few supernova, which also
provides the r-elements, should be the common formation mechanism.
And the $C_{r}$ scatter could be a natural result from the
different r-elements yields of supernova progenitors with
different masses. In addition, AIC or SN 1.5 should be the supplementary
scenario, especially for whose pre-AGB companion with higher mass
and smaller orbit radius, which support the both higher $C_{r}$
and $C_{s}$ values. Obviously, the fact should be that many origins of
r-elements for CEMP-r/s stars exist together.

Clearly, it is important for future studies to determine the
r-process sites and the precise r-process yields, which can
explain the abundance distribution of r-material in CEMP-r/s
stars. More in-depth theoretical and observational studies of
CEMP-s and CEMP-r/s stars will reveal more characteristics of the
production of r- and s-process elements at low metallicities, and
the history of enrichment of neutron-capture elements in our
Galaxy.

\acknowledgments
This work has been supported by the National Natural Science Foundation
of China under grant no.11003002, U1231119, 10673002, 10973016,
by the Natural Science Foundation
of Hebei Proince under Grant no. A2011205102, A2009000251,
by the Program for Excellent Innovative Talents in University of Hebei
Province under Grant no. CPRC034, by the Science Foundation of Hebei
Normal University under Grant no. L2009Z04, L2007B07.

\begin{table*}
 \centering
 \begin{minipage}{140mm}
\caption{The derived parameters
for the barium stars from \citet{all06a,pom08}.}
  \centering
  \begin{tabular}{@{}llccccrr@{}}
  \hline
 Star & [Fe/H] & $\Delta\tau$
&$r$ &$\tau_0$ &$C_s$ &$C_r$  &$\chi^2$
\\
& &(mb$^{-1}$) & &mb$^{-1}$ & & &\\
 \hline
 HR 107  &-0.36 &0.42 &0.10  &0.15
&0.0043 &0.9 &1.68534\\
 HD 749  &-0.06 &0.32 &0.18  &0.19
&0.0017 &0.8 &7.30354\\
 HD 5424  &-0.55 &0.52 &0.09  &0.22
&0.0006  &1.9  &10.52647\\
 HD 8270  &-0.42 &0.16 &0.26 &0.12
&0.0033  &1.7 &1.05882\\
 HD 12392 &-0.12 &0.52 &0.10 &0.22
&0.0003 &1.9 &10.52035\\
 HD 13551 &-0.24 &0.16 &0.25 &0.12
&0.0026 &0.7 &0.62892\\
 HD 22589 &-0.27 &0.42 &0.10 &0.18
&0.0006 &1.5 &1.64569\\
 HD 27271 &-0.09 &0.16 &0.25 &0.12
&0.0022 &1.8 &2.53701\\
 HD 48565 &-0.62 &0.33 &0.27 &0.25
&0.0014 &0.5 &2.41539\\
 HD 76225 &-0.31 &0.24 &0.12 &0.11
&0.0039 &1.2 &1.36991\\
 HD 87080 &-0.44 &0.30 &0.33 &0.27
&0.0008 &1.9 &4.31905\\
 HD 89948 &-0.30 &0.26 &0.10 &0.11
&0.0020 &1.1 &0.97308\\
 HD 92545 &-0.12 &0.44 &0.10 &0.14
&0.0004 &1.9 &1.07528\\
 HD 106191 &-0.29 &0.38 &0.10 &0.14
&0.0007 &1.3 &1.00532\\
 HD 107574 &-0.55 &0.48 &0.02 &0.12
&0.0004 &2.2 &4.76937\\
 HD 116869 &-0.32 &0.30 &0.32 &0.26
&0.0005 &0.9 &1.94790\\
 HD 123396 &-0.99 &0.56 &0.16 &0.31
&0.0003 &1.4 &13.98755\\
 HD 123585 &-0.48 &0.26 &0.28 &0.20
&0.0008 &4.3 &1.50356\\
 HD 147609 &-0.45 &0.28 &0.13 &0.14
&0.0018 &3.7 &2.43516\\
 HD 150862 &-0.10 &0.41 &0.10 &0.12
&0.0007 &1.4 &1.37586\\
 BD+18 5215 &-0.53 &0.46 &0.10 &0.10
&0.0012 &1.3 &2.24078\\
 HD 188985 &-0.30 &0.30 &0.17 &0.17
&0.0014 &1.1 &1.47145\\
 HD 210709 &-0.04 &0.32 &0.19 &0.19
&0.0004 &0.9 &3.30071\\
 HD 210910 &-0.37 &0.26 &0.17 &0.15
&0.0006 &3.3 &2.70146\\
 HD 222349 &-0.63 &0.31 &0.25 &0.22
&0.0015 &0.2 &1.28778\\
 HD 223938 &-0.13 &0.30 &0.25 &0.22
&0.0005 &1.0 &3.14901\\
 HD 11397 &-0.09 &0.16 &0.39 &0.17
&0.0011 &2.8 &0.47400\\
\hline
\end{tabular}
\end{minipage}
\end{table*}

\begin{table*}
 \centering
 \begin{minipage}{140mm}
  \caption{The derived parameters
for the barium stars from \citet{smi07}.
References:(a) Present work, (b) \citet{hus09}.}
  \centering
  \begin{tabular}{@{}llccccccrr@{}}
  \hline
 Star & [Fe/H] &[Pb/Fe]$^a$ &[Pb/Fe]$^b$ & $\Delta\tau$
&$r$ &$\tau_0$ &$C_s$ &$C_r$  &$\chi^2$
\\
& & & &(mb$^{-1}$) & &mb$^{-1}$ & & &\\
 \hline
 HD 46407 &-0.09 &0.84 &1.1 &0.27 &0.14 &0.14
&0.0024 &1.7 &4.02350\\
 HD 104979 &-0.35 &0.64 &0.6 &0.36 &0.16 &0.20
&0.0003 &1.3 &0.65889\\
 HD 116713 &-0.12 &0.50 &0.8 &0.27 &0.11 &0.12
&0.0016 &1.9 &3.15447\\
 HD 139195 &-0.02 &-0.24 &0.0 &0.31 &0.05 &0.10
&0.0004 &1.0 &0.23194\\
 HD 202109 &-0.04 &-0.2 &0.2 &0.25 &0.11 &0.11
&0.0004 &0.7 &0.25322\\
 HD 204075 &-0.09 &0.45 &1.2 &0.27 &0.06 &0.10
&0.0051 &0.4 &8.53757\\
 HD 205011 &-0.14 &0.03 &0.4 &0.25 &0.09 &0.10
&0.0013 &0.9 &0.86000\\
 HD 181053 &-0.19 &-0.04 &0.2 &0.28 &0.13 &0.14
&0.0003 &0.3 &0.17884\\
\hline
\end{tabular}
\end{minipage}
\end{table*}

\begin{figure*}
 \centering
 \includegraphics[width=1.6\textwidth,height=1.5\textheight]{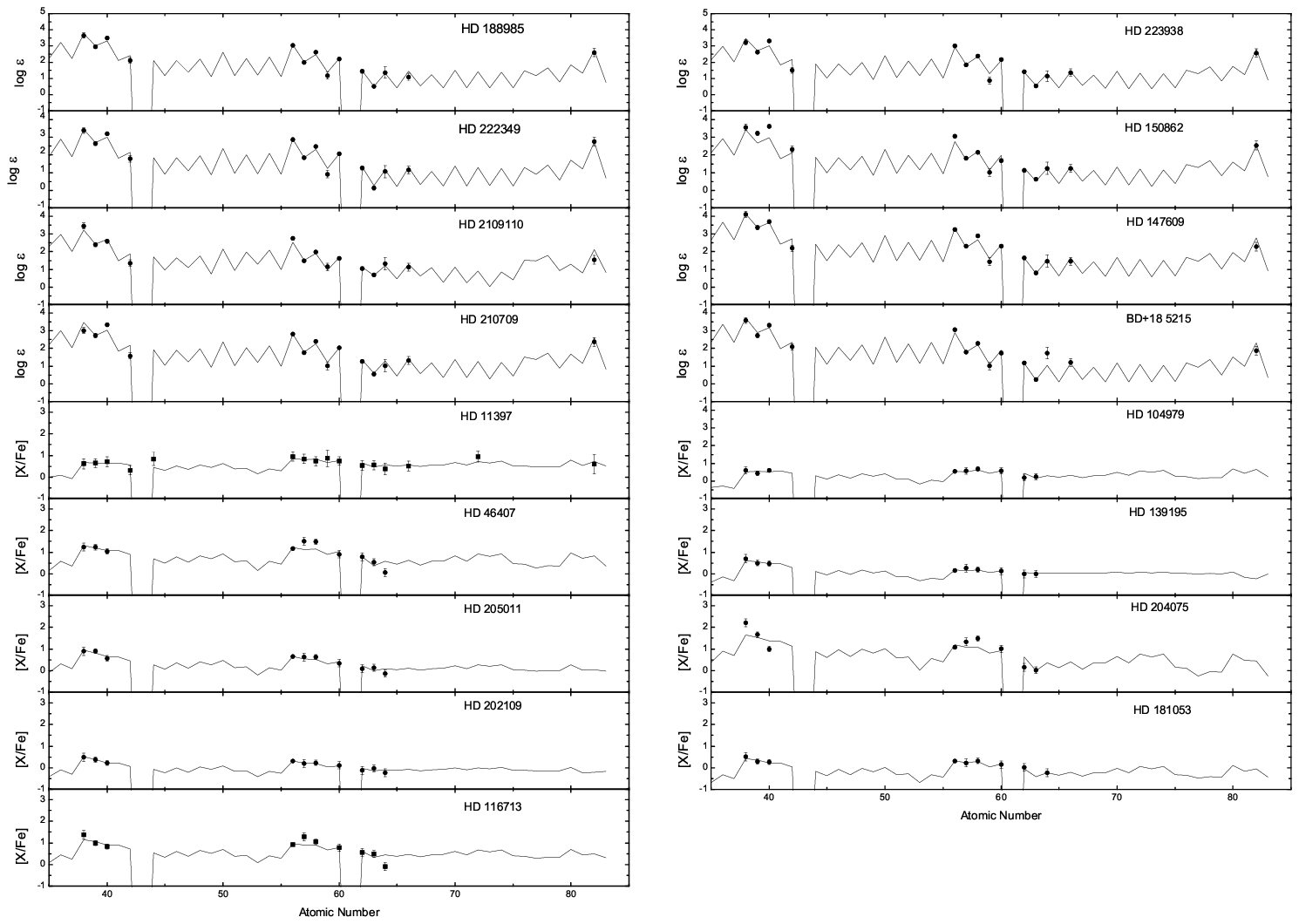}
 \suppressfloats[t]
 \renewcommand{\thefigure}{6}
 \vspace{-14.5cm}
 \caption{The rest part of figure~\ref{Fig:bafit}.}
\end{figure*}

\begin{table*}
 \centering
 \begin{minipage}{140mm}
  \caption{The derived parameters
for 24 s-rich metal-poor stars. References: (1) Present work,
(2) \citet{cui10}, (3) \citet{zha06}, (4) \citet{cui07a}, (5) \citet{cui07b}.}
  \centering
  \begin{tabular}{@{}lclccccrcc@{}}
  \hline
 Star &Class &[Fe/H] &$\Delta\tau$
&$r$ &$\tau_0$ &$C_s$ &$C_r$  &$\chi^2$ &ref
\\
& & &(mb$^{-1}$) & &mb$^{-1}$ & & & &\\
 \hline
 HE 0058-0244 &rs &-2.75 &0.72 &0.67  &1.80
&0.0014  &34.6  &2.39891 &(1)\\
 HE 0143-0441 &rs &-2.31 &0.68 &0.63 &1.47
&0.0035  &18.3 &2.19139 &(1)\\
 HE 1031-0020 &s &-2.86 &0.74 &0.73 &3.46
&0.0014 &4.6 &4.76720 &(1)\\
 HE 1509-0806 &s &-2.91 &0.23 &0.73 &0.73
&0.0031 &3.2 &0.88808 &(1)\\
 HE 2158-0348 &s &-2.70 &0.52 &0.58 &0.95
&0.0018 &3.8 &1.73201 &(1)\\
 HE 0024-2523 &s &-2.71 &0.70 &0.86 &4.64
&0.0022 &6.9 &4.23362 &(1)\\
 HE 0338-3945 &rs &-2.42 &0.76 &0.41 &0.85
&0.0049 &63.9 &1.00927 &(2)\\
 HE 2148-1247 &rs &-2.30 &0.88 &0.10 &0.38
&0.0045 &67.4 &2.36022 &(3)\\
 HE 1305-0007 &rs &-2.00 &0.71 &0.01 &0.15
&0.0047 &67.4 &1.290 &(4)\\
 CS 22183-015 &rs &-3.12 &0.66 &0.76  &2.40
&0.0028 &16.7 &2.40493 &(1)\\
 CS 22880-074 &s &-1.93 &0.60 &0.48  &0.82
&0.0005 &4.7 &1.82494 &(1)\\
 CS 22942-019 &s &-2.64 &0.43 &0.06 &0.15
&0.0041 &4.7 &2.26403 &(1)\\
 CS 31062-012 &rs &-2.55 &0.71 &0.32 &0.62
&0.0018 &37.3 &0.88371 &(3)\\
 CS 31062-050 &rs &-2.32 &0.71 &0.45 &0.89
&0.0039 &60.6 &1.02787 &(3)\\
 CS 29526-110 &rs &-2.38 &0.64 &0.79 &2.72
&0.0040 &50.6 &0.58090 &(3)\\
 CS 22898-027 &rs &-2.25 &0.77 &0.42 &0.89
&0.0035 &67.9 &1.22539 &(3)\\
 CS 22948-027 &rs &-2.47 &0.61 &0.37 &0.61
&0.0033 &65.9 &0.40286 &(3)\\
 CS 29497-034 &rs &-2.90 &0.53 &0.61 &1.07
&0.0034 &57.3 &1.10082 &(3)\\
 CS 29497-030 &rs &-2.57 &0.61 &0.81 &2.90
&0.0060 &86.4 &2.44198 &(3)\\
 CS 30301-015 &s &-2.64 &0.54 &0.34 &0.50
&0.0005 &0.9 &1.11864 &(3)\\
 CS 30322-023 &s &-3.41 &0.55 &0.65 &1.28
&0.0002 &0.08 &0.556 &(5)\\
 HD 196944 &s &-2.25 &0.45 &0.44 &1.26
&0.0006 &0.6 &0.58686 &(3)\\
 LP 625-44 &rs &-2.71 &0.69 &0.16 &0.38
&0.0045 &76.2 &2.11142 &(3)\\
 LP 706-7 &rs &-2.74 &0.82 &0.10 &0.36
&0.0017 &17.5 &0.84623 &(3)\\

\hline
\end{tabular}
\end{minipage}
\end{table*}

\end{document}